\documentclass[aps,prb,twocolumn,superscriptaddress,10pt]{revtex4-2}

\usepackage{amsmath, amssymb, mathtools}
\usepackage[version=4]{mhchem}
\usepackage{graphicx}
\usepackage{epsfig}
\usepackage{dcolumn}
\usepackage{bm}
\usepackage{blindtext}
\usepackage{natbib}
\usepackage{siunitx}
\usepackage{multirow}
\usepackage{bbm}
\usepackage[usenames,dvipsnames]{xcolor}
\usepackage{amsthm}
\usepackage{booktabs}
\usepackage{tikz}
\usetikzlibrary{calc}

\usepackage{hyperref}

\def\nn{\nonumber}

\newcommand{\ket}[1]{| #1 \rangle}
\newcommand{\bra}[1]{\langle #1 |}

\begin{document}

\title{Many-body wave function and edge magnetization  of an open $p+ is$ superconducting chain}

\author{Jiarui Jiao}
\affiliation{School of Physics, Nankai University, Tianjin, 300071, China}

\author{Chao Xu}
\affiliation{Institute for Advanced Study, Tsinghua University, Beijing 100084, China}
\affiliation{Kavli Institute for Theoretical Sciences, University of Chinese Academy of Sciences, Beijing 100190, China}

\author{Congjun Wu}
\email{wucongjun@westlake.edu.cn}
\affiliation{New Cornerstone Science Laboratory, Department of Physics, School of Science, Westlake University, Hangzhou 310024, Zhejiang, China}
\affiliation{Institute for Theoretical Sciences, Westlake University, Hangzhou 310024, Zhejiang, China}
\affiliation{Key Laboratory for Quantum Materials of Zhejiang Province, School of Science, Westlake University, Hangzhou 310024, Zhejiang, China}
\affiliation{Institute of Natural Sciences, Westlake Institute for Advanced Study, Hangzhou 310024, Zhejiang, China}

\author{Wang Yang}
\email{wyang@nankai.edu.cn}
\affiliation{School of Physics, Nankai University, Tianjin, 300071, China}

\begin{abstract}

Although BCS wave functions for superconductors under periodic boundary conditions are well established, 
obtaining an explicit form of the many-body BCS wave function under open boundary condition is usually a nontrivial problem.  
In this work, we construct the exact BCS ground-state wave function of a one-dimensional spin-$\frac12$ superconductor with $p+ is$  pairing symmetry under open boundary conditions for special sets of parameters. 
The spin magnetization on the edges is calculated explicitly using the obtained wave function. 
Our work is useful for obtaining deeper understandings of open $p+ is$ superconducting chains on a wave-function level. 

\end{abstract}

\maketitle

%%%%%%%%%%%%%%%%%%%%%%%%%%%%%%%%%%%%%%%%%%%%%%%%%%%%%%%%%%%
\section{Introduction}

Majorana fermions are particles that are their own antiparticles, which were originally proposed in high-energy physics \cite{Majorana1937} and later introduced into condensed matter systems as emergent quasiparticles \cite{Coleman1994}. 
Majorana zero modes  are localized, zero-energy excitations at the boundaries or defects of topological superconducting systems \cite{Kitaev2001,Tanaka2024,Niu2012,Fidkowski2011,Sato2016}.
Because of their global topological features \cite{Beenakker2020}, Majorana fermions and Majorana zero modes can encode quantum information nonlocally and save computational  resources  in fault-tolerant quantum computations \cite{Lian2018,O’Brien2018,You2014,Litinski2018,Kitaev2003,Sarma 2015}.  
Experimental signatures hinting the existences of Majorana zero modes have been reported in various condensed matter systems \cite{San-Jose2015,Jäck2019,Jack2021,Lutchyn2018,Sun2017,Haim2015,Liu2011}. 

Topological superconductors have attracted significant attentions due to their potential to host Majorana zero modes and Majorana fermions \cite{Niu2012,Fidkowski2011,Guo2016,Leijnse2012,Lin2017}. 
A well-established route to realizing Majorana zero modes is through $p$-wave topological superconductors. 
For instance, applying external magnetic fields to metallic nanowires in proximity with $s$-wave superconductors can drive the nanowire into a topological superconducting phase with effective spinless $p$-wave pairing, thereby creating the conditions necessary for the emergence of Majorana zero modes on the boundaries \cite{Kitaev2001,Qiao2022,Lesser2021,Bhullar2025}. 
Another example is Sr$_2$RuO$_4$, a two-dimensional superconductor that has been proposed as a candidate for unconventional chiral $p+ip$ pairing, where topologically protected Majorana zero modes  can be created in vortices \cite{Nelson2004,Mackenzie2017,Luke1998,Liu2015,Hughes2014}. 
More recently, the quasi-one-dimensional material K$_2$Cr$_3$As$_3$ has been reported to exhibit possible $p$-wave superconductivity and may be a promising platform for hosting Majorana zero modes \cite{Jiang2015,Yang2021}.

The $p\pm is$ superconducting state has recently attracted growing interest as an unconventional topological phase \cite{Gor'kov2001,Wu2010,Wang2011,Ryu2012,Stone2012,Qi2013,Wang2017,Yang2020,Roy2020,Sutradhar2024}. The $p\pm is$ pairing can be realized through two distinct approaches. One is an “extrinsic” route, in which an $s$-wave superconductor is placed in proximity to a $p$-wave superconductor, thereby inducing a mixed pairing state. The other is an “intrinsic” route, where the material itself possesses an inherent $p\pm is$ pairing symmetry. 
The $p\pm is$ pairing gives rise to several distinctive physical phenomena,
including axionlike electromagnetic response \cite{Goswami2014,Shiozaki2014,Stone2016,Xu2022} in which the relative phase between the $p$-wave and $s$-wave components acts as an effective axion field,
propagating chiral Majorana fermions along domain wall at the interface between adjacent $p+is$ and $p-is$ regions
and edge magnetizations \cite{Yang2020}. 

In this work, we focus on the one-dimensional (1D) $p+ is$ superconductors and demonstrate that at special parameters under open boundary conditions (OBC), the exact BCS many-body wave function for 1D $p+is$ superconductor can be obtained analytically,
exhibiting a recursive local form. 
We note that although the BCS wave function is well established under periodic boundary condition (PBC), 
obtaining the explicit form of the many-body BCS wave function under OBC is usually a nontrivial problem \cite{Greiter2014}.
The obtained results remain valuable even away from those special parameters, as long as  the system remains in the same phase. 

\begin{figure}[hbtp]
\includegraphics[width=0.5\textwidth]{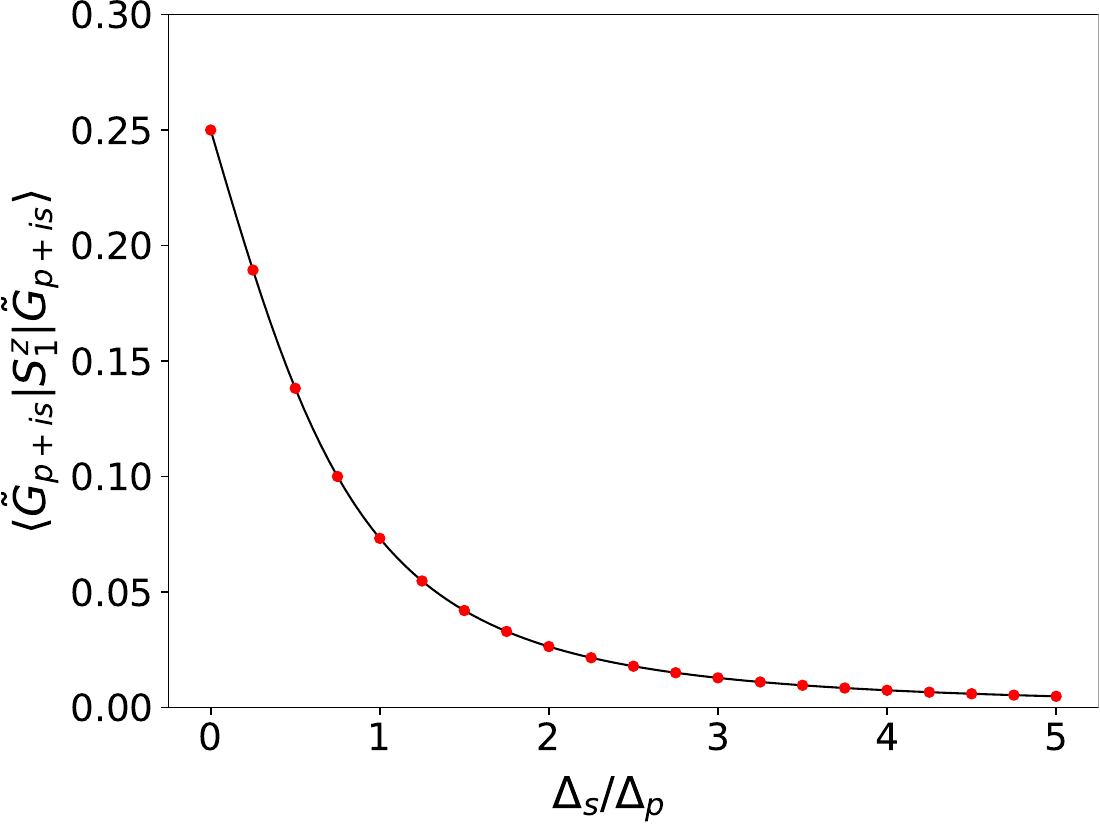}
\caption{
Numerical results for edge magnetization at site $1$ as a function of $\Delta_s/\Delta_p$  shown by the red dots, calculated on an open chain with $N=20$ sites.
Numerical values coincide with the analytical expression $\frac{1}{4}(1-\frac{\Delta_s/\Delta_p}{\sqrt{1+(\Delta_s/\Delta_p)^2}})$ plotted by the black curve.}
\label{fig:edgemag}
\end{figure}

The ground-state wave function under OBC can be useful for obtaining a better understanding for various physical properties of the $p+is$ superconducting  system on a wave function level. 
As an example, we further compute the spin magnetization on the edges based on the exact wave function,
and show that the result reported in Ref. \onlinecite{Yang2020} is not exact, but only an approximation valid when the $s$-wave component is much smaller than the $p$-wave one. 
More precisely, the edge magnetization has explicit dependence on the ratio between $s$- and $p$-wave components as shown in Fig. \ref{fig:edgemag},  
which reduces to the result of $\frac{1}{4}$magnetization in Ref. \onlinecite{Yang2020}  in the small $s$-wave limit. 
 
Apart from the calculation using wave function,  we present  another evaluation of edge magnetizations based on local Majorana operators. 
The special sets of system parameters which allow exact solutions of wave functions have a prominent feature that the spectrum of Bogoliubov eigen-modes is dispersionless. 
This allows the construction of scattering eigen-modes in terms of local operators. 
As a result, the contributions from boundary modes and scattering modes are both local in space,
providing a transparent understanding for the origins and separations  of these two types of contributions in edge magnetizations.

The rest of the paper is organized as follows. 
In Sec. \ref{sec:Ham} the model Hamiltonian for the spin-$1/2$ $p+is$ superconducting chain is discussed. 
Section \ref{sec:topo} shows that the 1D spin-$\frac12$ $p$-wave superconductor can be decomposed into two Kitaev superconducting chains with unpaired Majorana modes on the edges. 
In Sec. \ref{sec:exact}, the exact ground-state wave function is obtained in an explicit form in terms of creation operators. 
In Sec. \ref{sec:Edge magnetization}, the origin of edge spins in $p+is$ superconductors is explored through two approaches: calculation of edge magnetizations from the exact ground state wave function and reexamination within the picture of local Majorana modes.
Section \ref{sec:summary} summarizes the main results of the paper.  

%%%%%%%%%%%%%%%%%%%%%%%%%%%%%%%%%%%%%%%%%%%%%%%%%%%%%%%%%%%
\section{Model Hamiltonian}
\label{sec:Ham}

In this section, we briefly review the 1D $p+is$ superconductivity and edge magnetizations \cite{Yang2020}. 

%------------------------------------------------------------------------------------------------------------------------------------------------------------
\subsection{The 1D $p+is$ superconductor}

We first give a quick review on how the $p\pm is$ pairing state arises based on a free-energy analysis \cite{Yang2020}. 
The Ginzburg Landau free energy of the system with coexisting $p$- and $s$-wave superconducting pairing components up to quartic order can be written as
\begin{eqnarray}
F&=&\alpha_s|\Delta_s|^2+\alpha_p|\Delta_p|^2+\beta_s|\Delta_s|^4+\beta_p|\Delta_p|^4\nn\\
&&+\gamma|\Delta_s|^2|\Delta_p|^2+\eta\left(\Delta_s^*\Delta_s^*\Delta_p\Delta_p+\mathrm{c.c.}\right).
\label{eq:F}
\end{eqnarray}
It is clear that the free energy in Eq. (\ref{eq:F}) is invariant under both time reversal $\left(\mathcal T\right)$ and inversion $\left(\mathcal P\right)$ symmetries. 
The last term of the free energy $F$ in Eq. (\ref{eq:F}) can be rewritten as
\begin{flalign}
\eta(\Delta_s^*\Delta_s^*\Delta_p\Delta_p+\mathrm{c.c.})=2\eta |\Delta_s|^2|\Delta_p|^2\cos\left(2\phi_s-2\phi_p\right),
\end{flalign}
in which $\phi_s$ and $\phi_p$ are the phases of the complex order parameters $\Delta_s$ and $\Delta_p$, respectively,
whereas $|\Delta_s|$ and $|\Delta_p|$ are the amplitudes. 
For a positive $\eta$, minimizing the free energy leads to a difference of $\pm\pi/2$ between $\phi_s$ and $\phi_p$, giving rise to the $p\pm is$ pairing state.
In what follows throughout this work, we will focus on $p\pm is$ pairing, and choose $\Delta_p$ and $\Delta_s$ to be positive, representing the amplitudes of the pairing components.
In this convention, the value of the order parameter for the $s$-wave pairing is $\pm i \Delta_s$.

The 1D $p\pm is$ superconductor has been studied in Ref. \cite{Yang2020} using a continuum model. 
The Bogoliubov–de Gennes (BdG) Hamiltonian of the 1D spin-$1/2$ $p+is$ superconductor can be written as
\begin{flalign}
\hat{H}_{\mathrm{1D}}=\frac{1}{2}\int\mathrm dx\psi^\dagger\left(x\right)\left(\xi_{k_x}\sigma_3+\Delta_{k_x}\sigma_1\tau_1-\Delta_s\sigma_1\tau_2\right)\psi\left(x\right),
\label{1D}
\end{flalign}
in which 
$x$ is the spatial coordinate for the 1D system;
$\psi\left(x\right)=\left(c_\uparrow^\dagger\left(x\right)~c_\downarrow^\dagger\left(x\right)~c_\uparrow\left(x\right)~c_\downarrow\left(x\right)\right)^T$ is the Nambu spinor formed by electron creation and annihilation operators; 
$\sigma_i$ and $\tau_i$ are the Pauli matrices in the spin and particle-hole spaces, respectively; 
$\xi_{k_x}=\frac{\hbar^2}{2m}k_x^2-\mu\left(x\right)$ is the non interacting band dispersion; 
$\Delta_{k_x}=\frac{\Delta_p}{k_f}k_x$ is the pairing gap function for the $p$-wave superconducting component; 
$k_f$ is the Fermi wave vector; 
$k_x=-i\partial_x$ is the momentum operator; 
$\mu\left(x\right)$ is the chemical potential; 
$\Delta_s$ and $\Delta_p$ are positive-valued amplitudes of the $s$- and $p$-wave pairing amplitudes. 
The $\Delta_k$ and $\Delta_s$  terms in $\hat{H}_{\mathrm{1D}}$ in Eq. (\ref{1D}) correspond to the $p$ and $s$-wave pairing components, respectively.

When $\Delta_s=0$, the Hamiltonian reduces to that of a pure $p$-wave pairing superconductor.  
In this case, there are four Majorana zero modes in total, with two localized at each end of the system \cite{Yang2020}. 
As a result, the ground states are four fold degenerate. 
When $\Delta_s\neq0$, both inversion $\left(\mathcal P\right)$ and time-reversal $\left(\mathcal T\right)$ symmetries are broken,
though the system remains invariant under the combined $\mathcal P\mathcal T$ symmetry.
In this case, the four fold degeneracy of the Majorana zero modes is lifted and the ground state becomes non degenerate. 

%------------------------------------------------------------------------------------------------------------------------------------------------------------
\subsection{Spin magnetizations on the edges }

Since the system is 1D, there is no orbital angular momentum. 
However, it has been demonstrated that the 1D $p+is$ superconductors exhibit non vanishing spin magnetizations on the edges of an open chain \cite{Yang2020}. 

Expressing the spin operators $S^z_i$ in terms of the quasi particle operators,
the spin magnetizations in 1D $p+is$ superconductors can be calculated by evaluating the expectation values over the ground state of the system. 
In Ref. \onlinecite{Yang2020}, the quasi particle operators for the four low-energy Majorana modes are kept in the calculations, and it was found that the edge magnetizations are $\pm\frac14$, where the sign of the magnetization depends on the relative phase between $p$- and $s$-wave components as well as the location of the edge (i.e., left or right edge). 
Specifically, the magnetization at the same end of the chain in the $p-is$ case is opposite to that in the $p+is$ case;
and the two edges carry opposite magnetizations as ensured by the presence of the  $\mathcal P\mathcal T$ symmetry.

We note that although the treatments in Ref. \onlinecite{Yang2020} are standard and rather straightforward, 
the calculations are carried out based on the algebraic properties of the quasi particle operators, not touching the many-body wave functions. 
On the other hand, an explicit construction of the many-body wave-function of the 1D $p+is$ superconductor under open boundary condition can be helpful for obtaining a deeper and more intuitive understanding of the system. 
In particular, it will be desirable to explicitly see how edge magnetization emerges on a wave function level.

%%%%%%%%%%%%%%%%%%%%%%%%%%%%%%%%%%%%%%%%%%%%%%%%%%%%%%%%%%%
\section{The topological 1D $p$-wave superconductors}
\label{sec:topo}

In this section, we first briefly review Kitaev's spinless superconducting chain and the construction of its many-body wave function under open boundary condition.
After that, we show that the spin-$\frac12$ $p$-wave superconductor can be decomposed to two decoupled Kitaev's spinless superconducting chains.
As a result, the wave function of the spin-$\frac12$ $p$-wave superconductor under open boundary conditions can be obtained correspondingly. 

%------------------------------------------------------------------------------------------------------------------------------------------------------------
\subsection{Review of Kitaev superconducting chain}

We start with a brief review of Kitaev's model for spinless superconducting chain with unpaired Majorana zero modes on the edges \cite{Kitaev2001}. 

\subsubsection{Model Hamiltonian of Kitaev superconducting chain}

The model Hamiltonian for the Kitaev superconducting chain of $N$ sites under OBC is defined as
\begin{eqnarray}
\hat{H}_K&=&-t\sum_{x=1}^{N-1}\left(a_x^\dagger a_{x+1}+\text{h.c.}\right)-\Delta\sum_{x=1}^{N-1}\left(a_{x+1}^\dagger a_{x}^\dagger+\text{H.c.}\right)\nn\\
&&-\mu\sum_{x=1}^{N} a_x^\dagger a_x , 
\label{Kitaev}
\end{eqnarray}
in which $x$ ($1\leq x\leq N$) is the label for the lattice site; $N$ is the number of sites;
$a_x^\dagger$ and $a_x$ are creation and annihilation operators of a spinless fermion at site $x$, with no spin degrees of freedom;
$t$ is the nearest-neighbor hopping amplitude; $\Delta$ is the strength of the $p$-wave pairing strength; 
and $\mu$ is the chemical potential.

\subsubsection{Ground-state wave functions in real space}

Consider two special sets  of parameters of  Kitaev's superconducting model: $\mu=0,t>0,\Delta=\pm t$. 
At these two points, the wave function in real space for the ground states of the Hamiltonian in Eq. (\ref{Kitaev}) under OBC
 can be obtained analytically \cite{Greiter2014}, which we briefly review here. 
 
To begin with, we define two Majorana operators $\gamma_{A,x},\gamma_{B,x}$ on each site $x$ as
\begin{eqnarray}
\gamma_{A,x}&=&-i\left(a_x-a_x^\dagger\right),\nn\\
\gamma_{B,x}&=&a_x+a_x^\dagger. 
\end{eqnarray} 

For the case of $\Delta=t$, the Hamiltonian in Eq. (\ref{Kitaev}) can be rewritten as
\begin{eqnarray}
\hat{H}_{K,+}
&=&-it\sum_{x=1}^{N-1}\gamma_{B,x}\gamma_{A,x+1}.
\end{eqnarray}

The ground states of $\hat{H}_{K,+}$ are two fold degenerate and can be constructed explicitly \cite{Greiter2014}.
Denote $|0\rangle$ to be the vacuum state of the Fock space, namely, annihilated by the operators $a_x$ ($1\leq x\leq N$). 
The two degenerate ground states $|\Psi^\pm(\Delta=t)\rangle$ are given by
\begin{eqnarray}
|\Psi^\pm(\Delta=t)\rangle=\prod_{x=1}^N\frac{1}{\sqrt2}\left(1\pm a_x^\dagger\right)|0\rangle, 
\end{eqnarray}
in which the convention is chosen such that  the products act from right to left. For instance,
\begin{eqnarray}
\prod_{x=1}^N\left(1+a_x^\dagger\right)|0\rangle=\left(1+a_N^\dagger\right)\dots\left(1+a_1^\dagger\right)|0\rangle. 
\end{eqnarray}

Similarly, the Hamiltonian for $\Delta=-t$   is given by
\begin{eqnarray}
\hat{H}_{K,-}&=&-it\sum_{x=1}^{N-1}\gamma_{B,x+1}\gamma_{A,x}~.
\end{eqnarray}
In this case, the unpaired Majorana modes are $\gamma_{A,N}$ and $\gamma_{B,1}$.
It can be observed that the $\Delta=-t$ case is identical to the $\Delta=t$ case except that the ordering of the lattice sites is reversed. 
Therefore, the ground states $|\Psi^{\pm}(\Delta=-t)\rangle$ for $\Delta=-t$ can be obtained by by relabeling the sites in the ground states in the $\Delta=t$ case from $1,2,\dots,N$ to $N,N-1,\dots,1$.
More explicitly,
\begin{eqnarray}
|\Psi^{\pm}(\Delta=-t)\rangle=\prod_{x=N}^1\frac{1}{\sqrt2}\left(1\pm a_x^\dagger\right)|0\rangle, 
\end{eqnarray}

The Majorana operators $\gamma_{A,1}$ and $\gamma_{B,N}$ act as $-\sigma_y$ and $\sigma_z$ in the two-dimensional ground state subspace under the basis states $|\Psi^\pm(\Delta=t)\rangle$, while $\gamma_{A,N}$ and $\gamma_{B,1}$ act as $-\sigma_y$ and $\sigma_z$ under the basis states $|\Psi^\pm(\Delta=-t)\rangle$. These relations are explicitly verified in Appendix \ref{sec:I. The Summary of relations of operators}.  

%------------------------------------------------------------------------------------------------------------------------------------------------------------
\subsection{Spin-$\frac12$ $p$-wave superconductor as two decoupled Kitaev superconducting chains}
\label{eq:decoupled_K}

In this section, we rewrite the spin-$\frac12$ $p$-wave superconducting chain as two decoupled Kitaev chains with opposite pairing signs, which fixes the notation for subsequent constructions of wave functions under OBC.
The decoupling is a standard consequence of the Majorana representation and decomposition of free-fermion superconducting Hamiltonians, which have been systematically discussed in Refs. \onlinecite{Cobanera2015,Fu2021}.

\subsubsection{Decoupling of Hamiltonian under PBC}
\label{subsubsec:decouple}

The Bogoliubov–de Gennes (BdG) Hamiltonian of a 1D periodic $p$-wave pairing superconductor in the momentum space can be written as
\begin{eqnarray}
\hat{H}_{p}&=&\frac{1}{2}\sum_{k_x}\psi^\dagger\left(k_x\right)H_p\left(k_x\right)\psi\left(k_x\right),
\end{eqnarray}
in which the matrix $H_p(k_x)$ can be obtained by setting $\Delta_s$ as zero in Eq. (\ref{1D}),
namely,
\begin{eqnarray}
H_p\left(k_x\right)%&=&\xi_k\sigma_3+\Delta_k\sigma_1\tau_1\nn\\
&=&\begin{pmatrix}
\xi_{k_x}&0&0&\Delta_{k_x}\\
0&\xi_{k_x}&\Delta_{k_x}&0\\
0&\Delta_{k_x}&-\xi_{k_x}&0\\
\Delta_{k_x}&0&0&-\xi_{k_x}
\end{pmatrix}.
\label{BdG}
\end{eqnarray}
The operator vector $\psi\left(k_x\right)=(c_{k_x,\uparrow}~c_{k_x,\downarrow}~c_{-k_x,\uparrow}^\dagger~c_{-k_x,\downarrow}^\dagger)^T$  is related to $\psi^\dagger(x)$ in Eq. (\ref{1D}) via
\begin{eqnarray}
\psi^\dagger\left(k_x\right)=\frac{1}{\sqrt{N}}\sum_{x=1}^{N}\psi^\dagger(x) e^{-ik_xx},
\end{eqnarray}
where $k_x=\frac{2\pi n}{N}$ ($n=1,\dots,N$).

Consider the following unitary transformation on the electron creation and annihilation operators:
\begin{eqnarray} 
a_x&=&\frac{1}{\sqrt{2}}\left(c_{x,\uparrow}+c_{x,\downarrow}\right),\nn\\
a_x'&=&\frac{1}{\sqrt{2}}\left(c_{x,\uparrow}-c_{x,\downarrow}\right),
\label{eq:a_x_and_prime}
\end{eqnarray}
where $1\leq x\leq N$. Then the BdG Hamiltonian can be decomposed into a sum of two terms,
\begin{eqnarray}
\hat{H}_{p}&=&\hat{H}_{p,+}+\hat{H}_{p,-},
\end{eqnarray}
in which
\begin{eqnarray}
\hat{H}_{p,+}&=&\frac{1}{2}\sum_{k_x}\begin{pmatrix}a_{k_x}^\dagger&a_{-k_x}\end{pmatrix}
\begin{pmatrix}\xi_{k_x}&\Delta_{k_x}\\\Delta_{k_x}&-\xi_{k_x}\end{pmatrix}\begin{pmatrix}a_{k_x}\\a_{-k_x}^\dagger\end{pmatrix},\nn\\
\hat{H}_{p,-}&=&\frac{1}{2}\sum_{k_x}\begin{pmatrix}a_{k_x}'^\dagger&a_{-k_x}'\end{pmatrix}
\begin{pmatrix}\xi_{k_x}&-\Delta_{k_x}\\-\Delta_{k_x}&-\xi_{k_x}\end{pmatrix}\begin{pmatrix}a_{k_x}'\\a_{-k_x}'^\dagger\end{pmatrix}.\nn\\
\label{decompose}
\end{eqnarray}

Notice that up to the following gauge transformation 
\begin{eqnarray}
c_{j,\uparrow}&\rightarrow& e^{i\pi/4}c_{j,\uparrow},\nn\\
c_{j,\downarrow}&\rightarrow& e^{i\pi/4}c_{j,\downarrow},
\label{eq:gauge}
\end{eqnarray}
which is equivalent to 
$a_{k_x}\rightarrow e^{i\pi/4}a_{k_x}$, 
$a_{k_x}^\prime\rightarrow e^{i\pi/4}a_{k_x}^\prime$,
both $\hat{H}_{p,+}$ and $\hat{H}_{p,-}$ can be transformed to a spinless Kitaev  superconducting chain  in momentum space,
with opposite sign of the pairing. 
Therefore, the spin-$\frac12$ $p$-wave superconductor is equivalent to two copies of Kitaev  superconducting chains. 
We note that although $H_p\left(k_x\right)$ in Eq. (\ref{BdG}) can be decomposed into a direct sum of two $2\times 2$ matrices as well, 
the corresponding decomposition of $\hat{H}_{p}$ does not consist of two Kitaev superconducting chains,
since $c^\dagger_{j,\uparrow},c_{j,\downarrow}$ (similarly $c^\dagger_{j,\downarrow},c_{j,\uparrow}$) do not form a set of creation and annihilation operators for a single species of spinless fermion. 

\begin{figure}
\includegraphics[width=0.5\textwidth]{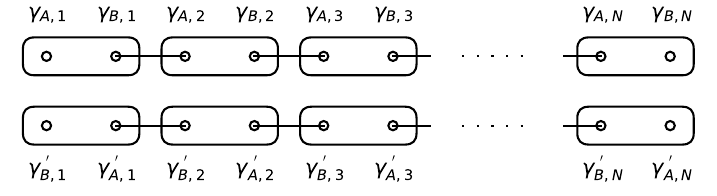}
\caption{Schematic plot of the pairing structure for the spin-${\frac12}$ $p$-wave superconducting chain at special parameters as two decoupled spinless Kitaev chains. 
The upper chain corresponds a copy of Kitaev superconducting chain with $\Delta=t$, described by the Hamiltonian $\hat{H}_{p,+}$ in Eq. (\ref{decompose}), while the lower chain corresponds to $\Delta=-t$, described by the Hamiltonian $\hat{H}_{p,-}$.}
\label{fig:chain3}
\end{figure}

Notice that $\hat{H}_{p,+}$ and $\hat{H}_{p,-}$ in Eq. (\ref{decompose}) differ in the sign of the off-diagonal pairing term $\pm\Delta_{k_x}$.
Hence, $\hat{H}_{p,+}$ ($\hat{H}_{p,-}$) describes a continuum model with $\Delta>0$ ($\Delta<0$).
The two decoupled Kitaev superconducting chains in Eq.~\ref{decompose} are both topologically nontrivial, albeit with opposite winding number $+1$ and $-1$,
and they lie in the same topological phases as those having special sets of parameters $(\mu=0,t>0,\Delta=t)$ and $(\mu=0,t>0,\Delta=-t)$, respectively.
The topological properties of the Kitaev model, including the winding number and the phase diagram, are detailed in Appendix \ref{sec:0. topo_properties}.
Since we are primarily interested in the global topological properties of the system, 
$\hat{H}_{p,+}$ (and $\hat{H}_{p,-}$) can be deformed into Kitaev superconducting chain with parameters $(\mu=0,t>0,\Delta=t)$ [and $(\mu=0,t>0,\Delta=-t)$] without affecting topological properties.
A schematic plot of the Hamiltonian $\hat H_p$ under OBC is shown in Fig. \ref{fig:chain3}. 
In what follows in this work, we will make the simplification that the  deformed Hamiltonian is considered. 

\subsubsection{Ground state under OBC}

Next we come back to OBC. 
After the above-mentioned deformation, the Hamiltonian $\hat{H}^{\text{D}}_p$ of an open spin-$\frac12$ $p$-wave superconducting chain becomes the sum of two Kitaev superconducting chains, with the two sets of parameters $(\mu=0,t>0,\Delta=t)$ and $(\mu=0,t>0,\Delta=-t)$,
where the superscript ``D" is ``deformed" for short. 
The Hamiltonian $\hat{H}^{\text{D}}_p$ under OBC can also be rewritten in terms of the Majorana operators as
\begin{eqnarray}
\hat{H}^{\text{D}}_p=-\frac{i\Delta_p}{2}\sum_{x=1}^{N-1}\left(\gamma_{B,x}\gamma_{A,x+1}+\gamma_{B,x+1}'\gamma_{A,x}'\right),
\label{eq:Hp_gamma}
\end{eqnarray}
in which $\gamma_{A,x},\gamma_{B,x}$ and $\gamma_{A,x}^\prime,\gamma_{B,x}^\prime$ are related to $c_{x,\uparrow},c_{x,\downarrow},c_{x,\uparrow}^\dagger,c_{x,\downarrow}^\dagger$ via 
\begin{eqnarray}
\gamma_{A,x}&=&-i\left(a_x-a_x^\dagger\right),\nn\\
\gamma_{B,x}&=&a_x+a_x^\dagger,\nn\\
\gamma_{A,x}^\prime&=&-i\left(a_x^\prime-a_x^{\prime\dagger}\right),\nn\\
\gamma_{B,x}^\prime&=&a^\prime_x+a_x^{\prime\dagger},
\label{eq:a_gamma}
\end{eqnarray}
where $a_x$ and $a_x^\prime$ are defined in Eq. (\ref{eq:a_x_and_prime}).

The ground-state wave function of $\hat{H}^{\text{D}}_p$ in Eq. (\ref{eq:Hp_gamma}) with OBC can be explicitly written. 
Since the ground states of each copy Kitaev superconducting chain are two  fold degenerate, the ground states  of $\hat{H}_p$ are four fold degenerate, denoted as $|\Psi^{++}\rangle,|\Psi^{+-}\rangle,|\Psi^{-+}\rangle,|\Psi^{--}\rangle$, which are given by
\begin{eqnarray}
|\Psi^{\pm\pm}\rangle=\prod_{x=1}^N\frac{1}{\sqrt2}\left(1\pm a_x^\dagger\right)\prod_{x=N}^1\frac{1}{\sqrt2}\left(1\pm a_x'^\dagger\right)|0\rangle.
\label{eq:Psi_pmpm}
\end{eqnarray}
The convention for the product is chosen as multiplying from right to left as before. 
The first  and second signs in the superscripts $\Psi$ correspond to the first and second products, respectively. 

%-------------------------------------------------------------------------

\begin{figure}
\includegraphics[width=0.5\textwidth]{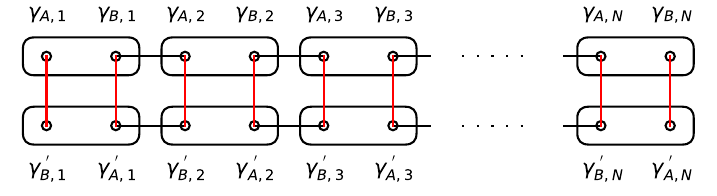}
\caption{Schematic plot of the pairing structure for the 1D  spin-$\frac{1}{2}$ $p+is$ superconducting model at special parameters. 
The interactions in the $p$ and $s$-wave pairing Hamiltonians are represented by the horizontal black lines and vertical red lines, respectively. }
\label{fig:chain}
\end{figure}
%-------------------------------------------------------------------------

%%%%%%%%%%%%%%%%%%%%%%%%%%%%%%%%%%%%%%%%%%%%%%%%%%%%%%%%%%%
\section{Exact wave function for $p+is$ superconducting chain}
\label{sec:exact}

In this section, we construct the exact ground-state wave function of an open spin-$\frac12$ $p+is$ superconducting chain.

\subsection{Bogoliubov transformation}
\subsubsection{Majorana representation of $p+is$ Hamiltonian}

Next we add $s$-wave component, considering a 1D $p\pm is$ superconductor with OBC. 
After performing the gauge transformation in Eq. (\ref{eq:gauge}), 
the  Hamiltonian $\hat{H}^{\text{D}}_{p+is}$ after deformation can be written as
\begin{flalign}
\hat{H}^{\text{D}}_{p+is}=-\frac{i\Delta_p}{2}\sum_{x=1}^{N-1}\left(\gamma_{B,x}\gamma_{A,x+1}+\gamma_{B,x+1}'\gamma_{A,x}'\right)+\hat{H}_s,
\label{eq:H_ps}
\end{flalign}
in which $\hat{H}_s$ is given by
\begin{eqnarray}
\hat{H}_s&=&\Delta_s\sum_{x=1}^N\left(c_{x,\uparrow}^\dagger c_{x,\downarrow}^\dagger+\text{H.c.}\right)\nn\\&=&\frac{i\Delta_s}{2}\sum_{x=1}^N\left(\gamma_{A,x}\gamma_{B,x}'+\gamma_{B,x}\gamma_{A,x}'\right).
\label{eq:Hs_gauge}
\end{eqnarray}
Here we emphasize that after the gauge transformation, the $p+is$ pairing becomes an $s+ip$ pairing, 
which is why there is no phase factor for $\Delta_s$ in Eq. (\ref{eq:Hs_gauge}). 

A pictorial representation of the Hamiltonian $\hat{H}^{\text{D}}_{p+is}$ in Eq. (\ref{eq:H_ps}) under OBC is shown in Fig. \ref{fig:chain}. 
In Fig. \ref{fig:chain}, the black and red lines represent the quadratic Majorana interactions in $\hat{H}^{\text{D}}_p$ and $\hat{H}_{s}$, respectively. 
Since $\hat{H}^{\text{D}}_p$ ($\hat{H}_{s}$) only contain intrachain (interchain) terms, the black (red) lines are correspondingly horizontal (vertical). 

\subsubsection{Orthogonal transformation of Majorana operators}

We recapitulate the Hamiltonian for the $p+is$ superconducting chain as follows,
\begin{eqnarray}
\hat{H}^{\text{D}}_{p+is}&=& \hat{H}_{\text{bulk}}+\frac{i\Delta_s}{2}\left(\gamma_{A,1}\gamma_{B,1}'+\gamma_{B,N}\gamma_{A,N}'\right),
\label{eq:H_ps_new}
\end{eqnarray}
in which $\hat{H}_{\text{bulk}}$ is given by
\begin{eqnarray}
\hat{H}_{\text{bulk}}&=&\sum_{x=1}^{N-1}\Gamma_{x+1/2}^T H_{x+1/2} \Gamma_{x+1/2},
\label{eq:H_ps_new_bulk}
\end{eqnarray}
where $H_{x+1/2}$ is a $4\times 4$ matrix given by
\begin{eqnarray}
H_{x+1/2}=-\frac{i}{4}\begin{pmatrix}
0&-\Delta_s&\Delta_p&0\\
\Delta_s&0&0&-\Delta_p\\
-\Delta_p&0&0&-\Delta_s\\
0&\Delta_p&\Delta_s&0
\end{pmatrix},
\end{eqnarray}
and $\Gamma_{x+1/2}$ is a four-component operator-valued column vector defined as
\begin{eqnarray}
\Gamma_{x+1/2}=(\gamma_{B,x}~\gamma^\prime_{A,x}~\gamma_{A,x+1}~\gamma^\prime_{B,x+1})^T.
\end{eqnarray}

By introducing a set of Bogoliubov-transformed operators $\tilde{\Gamma}_{x+1/2}=(\tilde{\gamma}_{B,x}~\tilde{\gamma}^\prime_{A,x}~\tilde{\gamma}_{A,x+1}~\tilde{\gamma}^\prime_{B,x+1})^T$ according to
\begin{eqnarray}
\tilde{\Gamma}_{x+1/2}^T=\Gamma_{x+1/2}^TO,
\label{eq:Bogoliubov}
\end{eqnarray}
in which $O$ is given by 
\begin{eqnarray}
O=\begin{pmatrix}
1&0&0&0\\
0& \frac{\Delta_p}{\sqrt{\Delta_s^2+\Delta_p^2}} & -\frac{\Delta_s}{\sqrt{\Delta_s^2+\Delta_p^2}} &0\\
0&\frac{\Delta_s}{\sqrt{\Delta_s^2+\Delta_p^2}} & \frac{\Delta_p}{\sqrt{\Delta_s^2+\Delta_p^2}} &0\\
0&0&0&1
\end{pmatrix},
\end{eqnarray}
the bulk Hamiltonian becomes
\begin{eqnarray}
\hat{H}_{\text{bulk}}&=&\sum_{x=1}^{N-1}\tilde{\Gamma}_{x+1/2}^T \tilde{H}_{x+1/2} \tilde{\Gamma}_{x+1/2},
\label{eq:standard_form}
\end{eqnarray}
in which 
\begin{eqnarray}
\tilde{H}_{x+1/2}=-\frac{i}{4}\sqrt{\Delta_s^2+\Delta_p^2}
\begin{pmatrix}
0&0&1&0\\
0&0&0&-1\\
-1&0&0&0\\
0&1&0&0
\end{pmatrix}.
\end{eqnarray}
Notice that  since the boundary Majorana modes are not touched by the transformation in Eq. (\ref{eq:Bogoliubov}),
the boundary term in Eq. (\ref{eq:H_ps_new}) remains unchanged, namely,
\begin{eqnarray}
\tilde\gamma_{A,1}&=&\gamma_{A,1},\nn\\
\tilde\gamma_{B,N}&=&\gamma_{B,N},\nn\\
\tilde\gamma_{B,1}'&=&\gamma_{B,1}',\nn\\
\tilde\gamma_{A,N}'&=&\gamma_{A,N}'.
\label{eq:boundary_untouched}
\end{eqnarray}

\begin{figure}
\includegraphics[width=0.5\textwidth]{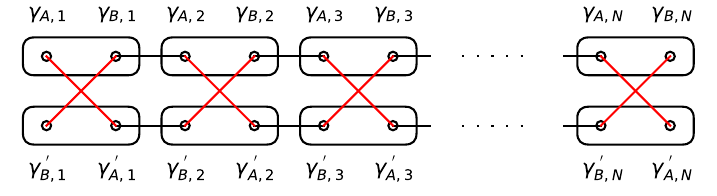}
\caption{Schematic plot of the pairing structure for the 1D spin-$\frac12$ $p+s$ superconducting model. The $s$-wave pairing, indicated by red lines, is nonlocal, in contrast to the $p+is$ case.}
\label{fig:chain2}
\end{figure}

We emphasize that it is the special structure of the $p+is$ pairing that enables a simple form of exact solution to be possible. 
As shown in Fig.~\ref{fig:chain2},
when there is no phase difference between $p$- and $s$-wave components, 
 the 1D $p+s$ superconducting model does not allow a simple reduction as Eq.~(\ref{eq:standard_form}), because the $s$-wave pairings between the two chains in the $p+s$ case are nonlocal. 

%---------------------------------------------------------------------------------------------------------------------------------------
%\subsection{Exact solution for ground state wave function}

\subsection{Decoupling to two chains}
\label{subsubsec:decouple_two_chains}

\begin{figure}
\centering
\includegraphics[width=0.5\textwidth]{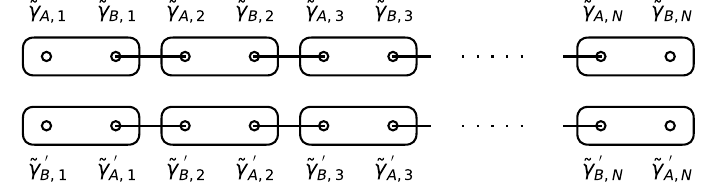}
\caption{Decoupling of spin-$\frac12$ $p+is$ superconductor into two spinless superconducting chains using $\tilde{\gamma}$-operators.}
\label{fig:chain3_1}
\end{figure}

The bulk Hamiltonian can be written in a more transparent form, as
\begin{flalign}
\hat{H}_{\text{bulk}}=-\frac{i}{2}\sqrt{\Delta_s^2+\Delta_p^2}\sum_{x=1}^{N-1}\left(\tilde\gamma_{B,x}\tilde\gamma_{A,x+1}+\tilde{\gamma}_{B,x+1}'\tilde{\gamma}_{A,x}'\right).
\label{eq:H_exact_bulk}
\end{flalign}
As shown in Fig. \ref{fig:chain3_1}, Eq. (\ref{eq:H_exact_bulk})  is of the same decoupled form as Fig. \ref{fig:chain3}, except that the $\gamma,\gamma^\prime$ operators have to be replaced with $\tilde{\gamma},\tilde{\gamma}^\prime$.  
Therefore, we can follow the same method in Sec. \ref{eq:decoupled_K} to construct the ground-state wave function. 

We introduce operators $\tilde{a}_x,\tilde{a}^\dagger_x,\tilde{a}^{\prime}_x,\tilde{a}^{\prime\dagger}_x$ ($1\leq x \leq N$) in a similar way as Eq. (\ref{eq:a_gamma}), i.e.,
\begin{eqnarray}
\tilde{\gamma}_{A,x}&=&-i\left(\tilde{a}_x-\tilde{a}_x^\dagger\right),\nn\\
\tilde{\gamma}_{B,x}&=&\tilde{a}_x+\tilde{a}_x^\dagger,\nn\\
\tilde{\gamma}_{A,x}^\prime&=&-i\left(\tilde{a}_x^\prime-\tilde{a}_x^{\prime\dagger}\right),\nn\\
\tilde{\gamma}_{B,x}^\prime&=&\tilde{a}^\prime_x+\tilde{a}_x^{\prime\dagger},
\label{eq:a_gamma_tilde}
\end{eqnarray}
in which the bulk and boundary Majorana operators on the left-hand side of the equations are given by Eqs. (\ref{eq:Bogoliubov}) and (\ref{eq:boundary_untouched}), respectively. 
We also introduce vacuum state $\ket{\tilde{0}}$, which is annihilated by $\tilde{a}_x$ and $\tilde{a}^\prime_x$ ($1\leq x \leq N$).
Notice that $\ket{\tilde{0}}$ can be constructed in the following way:
\begin{eqnarray}
\ket{\tilde{0}}=\frac{1}{\mathcal{N}}\Pi_{x=1}^N(\tilde{a}_x\tilde{a}_x^\prime)\ket{0},
\end{eqnarray}
in which $\mathcal{N}$ is a normalization factor. 
A closer inspection over the transformation in Eq. (\ref{eq:a_gamma_tilde}) reveals that $\tilde{a}_{x+1},\tilde{a}_{x+1}^\dagger,\tilde{a}_{x}^{\prime},\tilde{a}_{x}^{\prime\dagger}$ depend solely on $a_{x+1},a_{x+1}^\dagger,a_{x}^{\prime},a_{x}^{\prime\dagger}$.
As a result, it is convenient to rewrite $\ket{\tilde{0}}$ as 
\begin{eqnarray}
\ket{\tilde{0}}=\frac{1}{\mathcal{N}}\tilde{a}^{\prime}_N\tilde{a}_1\Pi_{x=1}^{N-1}(\tilde{a}_{x+1}\tilde{a}_x^\prime)\ket{0}.
\end{eqnarray}
Further calculations show that $\ket{\tilde{0}}$ can be expressed solely in terms of electron creation operators as 
\begin{eqnarray}
|\tilde0\rangle=\prod_{x=1}^{N-1}\left(u-v a_{x+1}^\dagger a_x'^\dagger\right)|0\rangle,
\label{eq:tilde0_cdagger}
\end{eqnarray}
in which 
\begin{eqnarray}
u&=&\sqrt{\frac12+\frac{\Delta_p}{2\sqrt{\Delta_s^2+\Delta_p^2}}}\;,\nn\\
v&=&\sqrt{\frac12-\frac{\Delta_p}{2\sqrt{\Delta_s^2+\Delta_p^2}}}\;.
\end{eqnarray}
Using Eq. (\ref{eq:a_x_and_prime}), it is clear that $|\tilde0\rangle$ in Eq. (\ref{eq:tilde0_cdagger}) has been expressed in the electron basis,  through the electron creation operators $c_{j,\uparrow}^\dagger,c_{j,\downarrow}^\dagger$. In Appendix \ref{sec:I. The Summary of relations of operators}, we explicitly verify the relations $\tilde{a}_{x}\ket{\tilde{0}}=\tilde{a}'_{x}\ket{\tilde{0}}=0$.

The ground states $|\tilde\Psi^{\pm\pm}\rangle$ of the bulk Hamiltonian $\hat{H}_{\text{bulk}}$ can be constructed in a way similar as Sec. \ref{eq:decoupled_K}, by replacing $a^\dagger_i,a^{\prime\dagger}_i$ with $\tilde{a}^\dagger_i,\tilde{a}^{\prime\dagger}_i$, yielding 
\begin{eqnarray}
|\tilde\Psi^{\pm\pm}\rangle=\prod_{x=1}^N\frac{1}{\sqrt2}\left(1\pm \tilde a_x^\dagger\right)\prod_{x=N}^1\frac{1}{\sqrt2}\left(1\pm \tilde a_x'^\dagger\right)|\tilde0\rangle.
\label{eq:G_exact_0}
\end{eqnarray}
The boundary terms in the full Hamiltonian in Eq. (\ref{eq:H_ps_new}) lift the ground-state degeneracy,  given by
\begin{eqnarray}
\hat{H}_{\text{edge}}&=&\frac{i\Delta_s}{2}\left(\gamma_{A,1}\gamma_{B,1}'+\gamma_{B,N}\gamma_{A,N}'\right),
\end{eqnarray}
Define two sets of Pauli matrices $\sigma_\alpha$, $\sigma^\prime_\beta$ ($\alpha,\beta=x,y,z$) acting on the four-dimensional  subspace spanned by $|\tilde\Psi^{\pm\pm}\rangle$, in accordance with
\begin{eqnarray}
\sigma_z |\tilde\Psi^{\lambda\mu}\rangle&=& \lambda |\tilde\Psi^{\lambda\mu}\rangle,\nn\\
\sigma^\prime_z |\tilde\Psi^{\lambda\mu}\rangle&=& \mu |\tilde\Psi^{\lambda\mu}\rangle,
\end{eqnarray}
where $\lambda,\mu=\pm$.
The actions of $\gamma_{A,1}$, $\gamma_{B,N}$, $\gamma^\prime_{A,N}$,and $\gamma^\prime_{B,1}$ can be derived as $-\sigma_y$, $\sigma_z$, 
$-\sigma_x\sigma^\prime_y$, and $\sigma_x\sigma^\prime_z$, respectively,
in which the $\sigma_x$ factors in $\gamma_{A,N}$ and $\gamma_{B,1}$ are due to sign flips in $(1\pm\tilde a_x^\dagger)$'s when commuting $\gamma^\prime_{A,N},\gamma^\prime_{B,1}$ through $\prod_{x=1}^N\left(1\pm\tilde a_x^\dagger\right)$.
As a result, the restriction of $\hat{H}_{\text{edge}}$ in the four-dimensional subspace spanned by $|\tilde\Psi^{\pm\pm}\rangle$ becomes a $4\times 4$ matrix $H_{\text{res}}$, given by 
\begin{eqnarray}
H_{\text{res}}=\frac{\Delta_s}{2} (\sigma_y\sigma^\prime_y-\sigma_z\sigma^\prime_z).
\label{matrix}
\end{eqnarray}
The ground state of $H_{s,\text{res}}$ can be easily solved as $\frac{1}{\sqrt{2}}(1,0,0,1)^T$, with energy $-\frac{\Delta_s}{2}$.
Hence, the exact ground state of the 1D $p+is$ superconductor under OBC is given by 

\begin{eqnarray}
|\tilde G_{p+is}\rangle=\frac{1}{\sqrt{2}}\left(|\tilde\Psi^{++}\rangle+|\tilde\Psi^{--}\rangle\right).
\label{eq:G_exact_1}
\end{eqnarray}

\subsection{Exact wave function}

However, since $\tilde a_x^\dagger$ and $\tilde a_x^{\prime\dagger}$ contain both electron creation and annihilation operators, the expression of the ground state in Eq. (\ref{eq:G_exact_1}) is still not of the desired form for the many-body wave function.
We still need to eliminate the electron annihilation operators from Eq. (\ref{eq:G_exact_1}).

As a first step, we move operators with the same site index together, which can be achieved in a recursive way as discussed in details in Appendix \ref{sec:III. The recursive form of the exact ground state wave function}.
Here we briefly sketch how the recursion is carried out. 
We introduce the notations $\tilde{\phi}_{n,m}^{\lambda},\tilde{\phi}_{m,n}^{\prime\lambda}$ defined as
\begin{eqnarray}
\tilde{\phi}_{n,m}^{\lambda}&=& \frac{1}{\sqrt{2}^{n-m+1}}(1+\lambda\tilde{a}^\dagger_n)(1+\lambda\tilde{a}^\dagger_{n-1})...(1+\lambda\tilde{a}^\dagger_m),\nn\\
\tilde{\phi}_{m,n}^{\prime\lambda}&=& \frac{1}{\sqrt{2}^{n-m+1}}(1+\lambda\tilde{a}^{\prime\dagger}_m)(1+\lambda\tilde{a}^{\prime\dagger}_{m+1})...(1+\lambda\tilde{a}^{\prime\dagger}_n),\nn\\
\end{eqnarray}
in which $\lambda=\pm$ and $1\leq m<n\leq  N$. 
It can be shown that $|\tilde \Psi^{++}\rangle$ can be written  as
\begin{flalign}
|\tilde \Psi^{++}\rangle=\frac{1}{2^m} \big(
\tilde{\phi}_{m,1}^{+}\tilde{\phi}_{1,m}^{\prime+} \hat{h}_m^{++}|\tilde0\rangle+\tilde{\phi}_{m,1}^{-}\tilde{\phi}_{1,m}^{\prime-} \hat{h}_m^{--}|\tilde0\rangle
\big),
\end{flalign}
in which $\hat{h}_m^{++}$ and $\hat{h}_m^{--}$ can be determined recursively according to
\begin{flalign}
\begin{pmatrix}\hat h_{m-1}^{++}\\\hat h_{m-1}^{--}\end{pmatrix}=
\begin{pmatrix}
1+\tilde a_m'^\dagger&-\tilde a_m^\dagger\left(1-\tilde a_m'^\dagger\right)\\
\tilde a_m^\dagger\left(1+\tilde a_m'^\dagger\right)&1-\tilde a_m'^\dagger
\end{pmatrix}
\begin{pmatrix}\hat h_{m}^{++}\\\hat h_{m}^{--}\end{pmatrix},
\label{eq:move_together_tilde}
\end{flalign}
where the recursion starts at $m=N$ with $\hat{h}_{N}^{++} = 1$, $\hat{h}_{N}^{--} = 0$,
and proceeds with decreasing $m$.
As a result,  $|\tilde \Psi^{++}\rangle$ can be expressed as
\begin{flalign}
&|\tilde \Psi^{++}\rangle\nn\\
&=\frac{1}{2^N}\tilde{\eta}_1^T\prod_{x=N}^2\begin{pmatrix}
1+\tilde a_x'^\dagger&-\tilde a_x^\dagger\left(1-\tilde a_x'^\dagger\right)\\
\tilde a_x^\dagger\left(1+\tilde a_x'^\dagger\right)&1-\tilde a_x'^\dagger
\end{pmatrix}f_N^{++}|\tilde0\rangle,
\label{eq:recurs}
\end{flalign}
where the column vectors  are defined as
\begin{eqnarray}
\tilde{\eta}_1&=&\begin{pmatrix}(1+\tilde a_1^\dagger)(1+\tilde a_1'^\dagger),(1-\tilde a_1^\dagger)(1-\tilde a_1'^\dagger)\end{pmatrix}^T,\nn\\
f_N^{++}&=&\begin{pmatrix}1,0\end{pmatrix}^T.
\end{eqnarray} 
For $|\tilde \Psi^{--}\rangle$ , it suffices to replace $f_N^{++}$ with $f_N^{--}$, where $f_N^{--}=\begin{pmatrix}0,1\end{pmatrix}^T$ .

Next, the electron annihilation operators need to be removed from $|\tilde \Psi^{++}\rangle$ and $|\tilde \Psi^{--}\rangle$. 
Notice that $|\tilde0\rangle$ in Eq. (\ref{eq:recurs}) contains factors as $\left(u-v a_{m}^\dagger a_{m-1}'^\dagger\right)$, 
hence, the following formulas for the actions of the operators $\{1,\tilde a_{m-1}'^\dagger,\tilde a_{m}^\dagger,\tilde a_{m}^\dagger\tilde a_{m-1}'^\dagger\}$ on $\left(u-v a_{m}^\dagger a_{m-1}'^\dagger\right)$ are useful for further reduction 
\begin{eqnarray}
&&\begin{pmatrix}1,\tilde a_{m-1}'^\dagger,\tilde a_{m}^\dagger,\tilde a_{m}^\dagger\tilde a_{m-1}'^\dagger\end{pmatrix}\left(u-v a_{m}^\dagger a_{m-1}'^\dagger\right)\nn\\
&=&\begin{pmatrix}1,a_{m-1}'^\dagger,a_{m}^\dagger,a_{m}^\dagger a_{m-1}'^\dagger\end{pmatrix}
\begin{pmatrix}
u&0&0&v\\
0&u&-v&0\\
0&v&u&0\\
-v&0&0&u
\end{pmatrix},
\label{eq:action}
\end{eqnarray}
which eliminates the electron annihilation operators $a_m,a_m^\prime$ from the operators $\tilde{a}_m,\tilde{a}_m^\prime$ for Bogoliubov quasiparticles. 
In what follows, we briefly sketch how the removal of electron annihilation operators is carried out in the expression of $|\tilde \Psi^{++}\rangle$, with details included in Appendix \ref{sec:III. The recursive form of the exact ground state wave function}.

Since $u-v a_{m}^\dagger a_{m-1}'^\dagger$ is a bosonic operator, it can be freely moved around. 
Then $|\tilde \Psi^{++}\rangle$ can be rewritten as
\begin{flalign}
&|\tilde \Psi^{++}\rangle=\frac{1}{2^N}\tilde{\eta}_1^T(u-v a_{2}^\dagger a_{1}'^\dagger)\nn\\
&\times\left[\prod_{x=m-1}^2\begin{pmatrix}
1+\tilde a_x'^\dagger&-\tilde a_x^\dagger\left(1-\tilde a_x'^\dagger\right)\\
\tilde a_x^\dagger\left(1+\tilde a_x'^\dagger\right)&1-\tilde a_x'^\dagger
\end{pmatrix} (u- v a_{x+1}^\dagger a_{x}'^\dagger)\right]\nn\\
&\times \hat{g}_m^{++}\ket{0}.
\label{eq:remove_a_Psi_pp}
\end{flalign}
In Eq. (\ref{eq:remove_a_Psi_pp}), the operator $\hat{g}_m^{++}$ ($2\leq m\leq N$) can be shown to exhibit the form
\begin{eqnarray}
\hat{g}_m^{++}=\begin{pmatrix}J_{m}-\tilde a_{m}^\dagger K_{m}\\K_{m}+\tilde a_{m}^\dagger J_{m}\end{pmatrix},
\end{eqnarray}
in which $J_{m}$ and $K_{m}$ can be determined recursively as
\begin{eqnarray}
J_{m-1}&=&\Lambda_{11}^{m}J_m+\Lambda_{12}^{m}K_m,\nn\\
K_{m-1}&=&\Lambda_{21}^{m}J_m+\Lambda_{22}^{m}K_m,
\end{eqnarray}
where the constant operators $\Lambda_{ij}^{m}$ ($i,j=1,2$) are given by 
\begin{eqnarray}
\Lambda_{11}^{m}&=&u+ua_{m-1}'^\dagger+va_{m}^\dagger-va_{m}^\dagger a_{m-1}'^\dagger,\nn\\
\Lambda_{12}^{m}&=&v+va_{m-1}'^\dagger-ua_{m}^\dagger+ua_{m}^\dagger a_{m-1}'^\dagger,\nn\\
\Lambda_{21}^{m}&=&v-va_{m-1}'^\dagger+ua_{m}^\dagger+ua_{m}^\dagger a_{m-1}'^\dagger,\nn\\
\Lambda_{22}^{m}&=&u-ua_{m-1}'^\dagger-va_{m}^\dagger-va_{m}^\dagger a_{m-1}'^\dagger.
\label{eq:Lambda_ijm}
\end{eqnarray}
The recursion starts at $m=N$ with $J_N,K_N$ given by 
\begin{eqnarray}
J_N&=&1+a_N'^\dagger,\nn\\
K_N&=&1-a_N'^\dagger,
\end{eqnarray}
and proceeds by decreasing $m$ until $m=2$.
The expression for $|\tilde \Psi^{--}\rangle$ can be obtained in an exactly similar recursive manner.

After obtaining the final operators $J_1,K_1$ by the end of the recursion, and combining $|\tilde \Psi^{++}\rangle,|\tilde \Psi^{--}\rangle$, the ground state $|\tilde G_{p+is}\rangle$ can be simplified to
\begin{eqnarray}
|\tilde G_{p+is}\rangle&=&\frac{1}{\sqrt{2}^{2N+1}}\left[\left(1+a_1^\dagger\right)J_1+\left(1-a_1^\dagger\right)K_1\right]|0\rangle\nn\\
&=&\frac{1}{\sqrt{2}^{2N+1}}\eta_1^T\left[\prod_{x=N}^2\begin{pmatrix}\Lambda_{11}^{x}&\Lambda_{12}^{x}\\\Lambda_{21}^{x}&\Lambda_{22}^{x}
\end{pmatrix}\right]\begin{pmatrix}1+a_N'^\dagger\\1- a_N'^\dagger\end{pmatrix}|0\rangle,\nn\\
\label{eq:exact_wf}
\end{eqnarray}
in which $\Lambda_{ij}^m$ ($i,j=1,2$) are given in Eq. (\ref{eq:Lambda_ijm}), and $\eta_1 $ is defined as
\begin{eqnarray}
\eta_1^T=(1+ a_1^\dagger,1-a_1^\dagger).
\end{eqnarray} 
Equation (\ref{eq:exact_wf}) is the desired wave function for $p+is$ superconductor under OBC,
which only contains electron creation operators. 

We note that the exact many-body wave function obtained in Eq. (\ref{eq:exact_wf}) appears to exhibit a highly organized internal structure. 
Its recursive local form suggests that it may offer a useful starting point for future studies of the entanglement properties of the open-chain ground state, for possible matrix-product-state/tensor-network descriptions, and for studying local measurements. 
In addition, the local connectivity pattern shown in Fig. 4 suggests a possible relation to a fixed-point representative wave function of the underlying phase, in the sense that a simple block decimation appears to preserve the basic form of the couplings to the surrounding degrees of freedom and hence indicates a degree of self-similarity under coarse graining. 
We leave a systematic exploration of these issues to future work.

%%%%%%%%%%%%%%%%%%%%%%%%%%%%%%%%%%%%%%%%%%%%%%%%%%%%%%%%%%%
\section{Edge magnetization}
\label{sec:Edge magnetization}

In this section, we calculate the edge magnetizations in $p+is$ superconducting chains using the exact ground-state wave function obtained in Sec. \ref{sec:exact}.
We also present an alternative calculation of edge magnetizations based on the local Majorana modes in Fig. \ref{fig:chain}.

\subsection{Wave-function calculation of edge magnetization}

Having obtained the exact ground-state wave function, we next evaluate the edge magnetization on a wave-function level using the form in Eq. (\ref{eq:exact_wf}).

We introduce the notations $\varphi_{1,m}$ and $|\varphi_{m,N}\rangle$ defined as
\begin{eqnarray}
\varphi_{1,m}&=&\frac{1}{\sqrt{2}^{2m-1}}\eta_1^T\left[\prod_{x=m}^{1}\begin{pmatrix}\Lambda_{11}^{x}&\Lambda_{12}^{x}\\\Lambda_{21}^{x}&\Lambda_{22}^{x}\end{pmatrix}\right]\nn\\
|\varphi_{m,N}\rangle&=&\frac{1}{\sqrt{2}^{2N-2m+2}}\left[\prod_{x=N}^{m+1}\begin{pmatrix}\Lambda_{11}^{x}&\Lambda_{12}^{x}\\\Lambda_{21}^{x}&\Lambda_{22}^{x}
\end{pmatrix}\right]\begin{pmatrix}1+a_N'^\dagger\\1- a_N'^\dagger\end{pmatrix}|0\rangle.\nn\\
\end{eqnarray}
For the magnetization at the site $1$, the ground state $|\tilde G_{p+is}\rangle$ can be expressed as
\begin{eqnarray}
|\tilde G_{p+is}\rangle&=&\frac{1}{\sqrt{2}^{3}}\eta_1^T\begin{pmatrix}\Lambda_{11}^{2}&\Lambda_{12}^{2}\\\Lambda_{21}^{2}&\Lambda_{22}^{2}
\end{pmatrix}|\varphi_{2,N}\rangle.
\end{eqnarray}
Then the magnetization of the ground state $|\tilde G_{p+is}\rangle$ at site $1$ is 
\begin{eqnarray}
&&\langle \tilde G_{p+is}|S_1^z|\tilde G_{p+is}\rangle\nn\\ &=&\frac18\langle\varphi_{2,N}|\begin{pmatrix}\Lambda_{11}^{2\dagger}&\Lambda_{21}^{2\dagger}\\\Lambda_{12}^{2\dagger}&\Lambda_{22}^{2\dagger}
\end{pmatrix}\eta_1^{T\dagger}S_1^z\eta_1^T\begin{pmatrix}\Lambda_{11}^{2}&\Lambda_{12}^{2}\\\Lambda_{21}^{2}&\Lambda_{22}^{2}
\end{pmatrix}|\varphi_{2,N}\rangle\nn\\
&=&\frac18\langle\varphi_{2,N}|\begin{pmatrix}2(u-v)^2&0\\0&2(u-v)^2
\end{pmatrix}|\varphi_{2,N}\rangle\nn\\
&=&\frac14(u-v)^2=\frac{1}{4}\left(1-\frac{\Delta_s}{\sqrt{\Delta_s^2+\Delta_p^2}}\right).
\label{eq:magnetization_exact}
\end{eqnarray}
Notice that Eq. (\ref{eq:magnetization_exact}) reduces to $\frac14$ for small $\Delta_s$,
and approaches $0$ when $\Delta_s\gg \Delta_p$.

A similar calculation can also derive the magnetization at site $N$.
For the magnetization at the site $N$, the ground state $|\tilde G_{p+is}\rangle$ can be expressed as
\begin{eqnarray}
|\tilde G_{p+is}\rangle=\frac{1}{\sqrt{2}^4}\varphi_{1,N-1}\begin{pmatrix}\Lambda_{11}^{N}&\Lambda_{12}^{N}\\\Lambda_{21}^{N}&\Lambda_{22}^{N}\end{pmatrix}\begin{pmatrix}1+a_N'^\dagger\\1- a_N'^\dagger\end{pmatrix}|0\rangle.
\end{eqnarray}
The magnetization of the ground state $|\tilde G_{p+is}\rangle$ at site $N$ is given by
\begin{eqnarray}
&&\langle \tilde G_{p+is}|S_N^z|\tilde G_{p+is}\rangle\nn\\ &=&\frac{1}{16}\langle0|\begin{pmatrix}1+a_N'&1- a_N'\end{pmatrix}\begin{pmatrix}\Lambda_{11}^{N\dagger}&\Lambda_{21}^{N\dagger}\\\Lambda_{12}^{N\dagger}&\Lambda_{22}^{N\dagger}\end{pmatrix}\nn\\
&&\times\varphi_{1,N-1}^\dagger S_N^z\varphi_{1,N-1}\begin{pmatrix}\Lambda_{11}^{N}&\Lambda_{12}^{N}\\\Lambda_{21}^{N}&\Lambda_{22}^{N}
\end{pmatrix}\begin{pmatrix}1+a_N'^\dagger\\1- a_N'^\dagger\end{pmatrix}|0\rangle\nn\\
&=&\frac{1}{16}\langle0|-4(u-v)^2|0\rangle=-\frac14(u-v)^2.
\label{eq:magnetization_exact2}
\end{eqnarray}
The result shows that the magnetization at site $N$ is $-\frac{1}{4}(1-\frac{\Delta_s}{\sqrt{\Delta_s^2+\Delta_p^2}})$,
which is opposite to the magnetization at site $1$, consistent with the $\mathcal{PT}$-symmetry of the system. 
We note that one can also evaluate the magnetization $\langle \tilde G_{p+is}|S_N^z|\tilde G_{p+is}\rangle$ by directly calculating the expectation value using Eqs. (\ref{eq:G_exact_0}) and  (\ref{eq:G_exact_1}).

\begin{figure}
\includegraphics[width=0.45\textwidth]{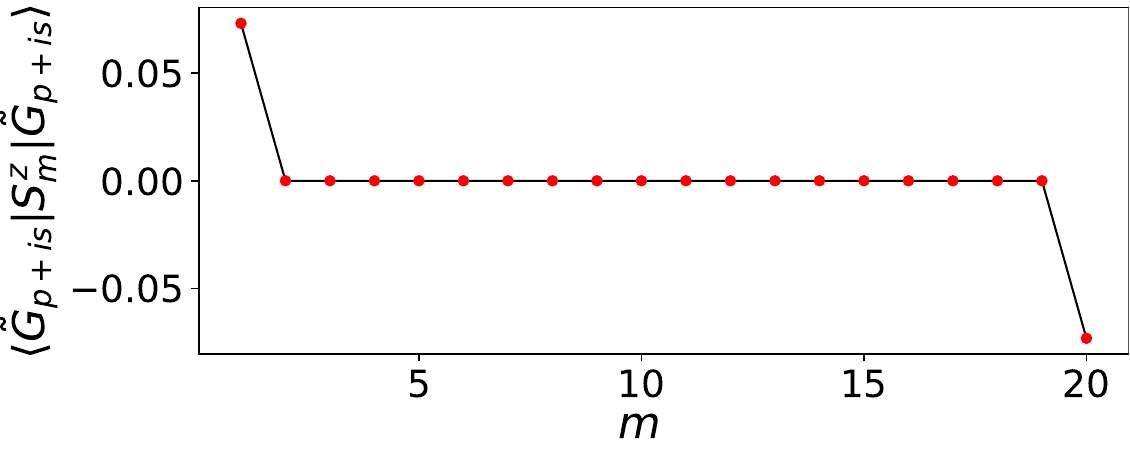}
\caption{Site-resolved magnetization $\bra{\tilde{G}_{p+is}} S_m^z \ket{\tilde{G}_{p+is}}$ as a function of site $m$ at $\mu=0$, $\Delta_p = \Delta_s=t$ on an open $p+is$ superconducting chain with $N = 20$ sites. 
The magnetization  in the bulk vanishes, while its values at the edge sites $1$ and $N$ agree well with the analytical results, which are $\pm \frac{1}{4}\left(1-\frac{1}{\sqrt{2}}\right)\approx \pm 0.073$.}
\label{fig:edge}
\end{figure}

\begin{figure}
\includegraphics[width=0.45\textwidth]{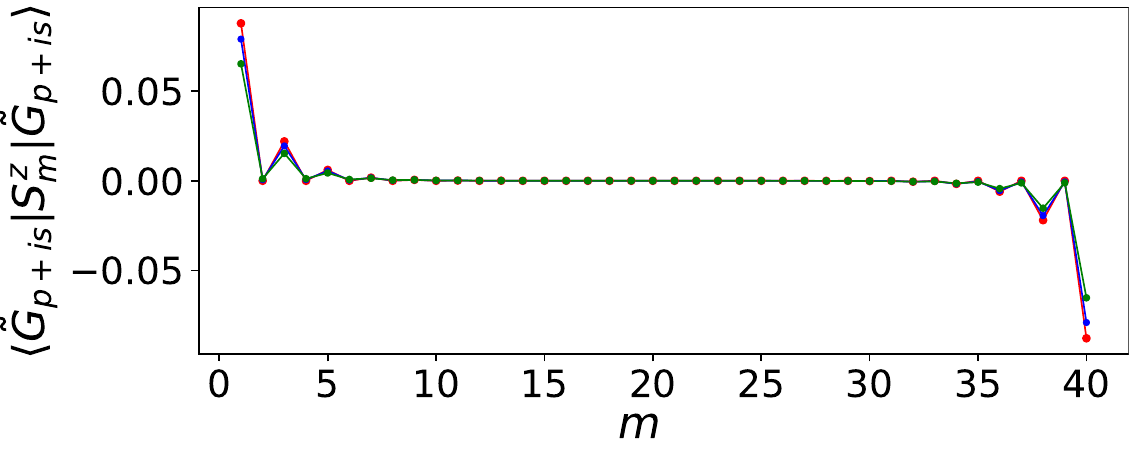}
\caption{Site-resolved magnetization $\bra{\tilde{G}_{p+is}} S_m^z \ket{\tilde{G}_{p+is}}$ as a function of site $m$ at $\Delta_p =5t$, $\Delta_s=t$ in the $p+is$ superconductor with $N=40$ sites, where 
the chemical potential is chosen as $\mu=0$ (red), $\mu=t$ (blue), and $\mu=1.7t$ (green).
The magnetization extends over several sites near the edge and decays to zero toward the bulk.}
\label{fig:variety}
\end{figure}

\begin{figure}
\includegraphics[width=0.45\textwidth]{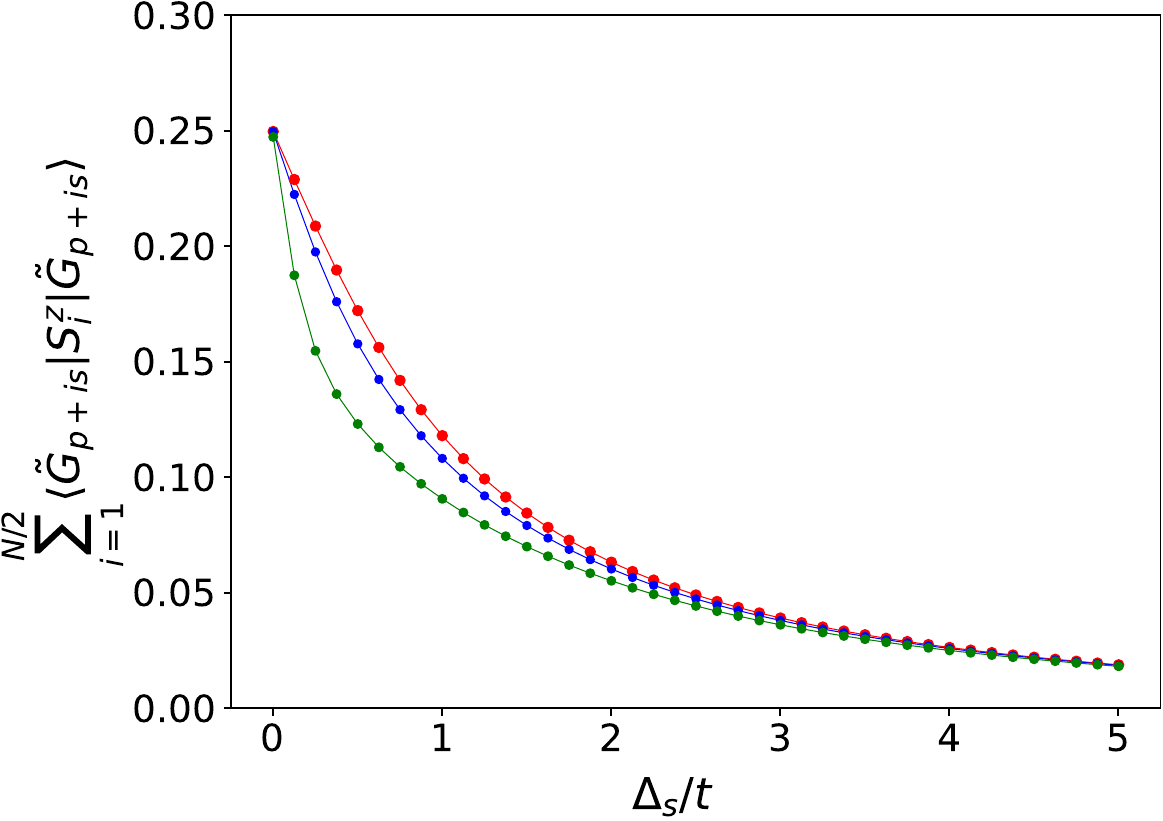}
\caption{Magnetization $\sum_{i=1}^{N/2}\bra{\tilde{G}_{p+is}} S_i^z \ket{\tilde{G}_{p+is}}$ summed over half of the chain for $\mu=0$ (red), $\mu=t$ (blue) and $\mu=1.7t$ (green) with $\Delta_p=5t$ and $N=240$ sites. 
The summed magnetization evolves from $+\frac14$ to $0$ with increasing strength of $\Delta_s$.}
\label{fig:varymag}
\end{figure}

For the magnetization in the bulk at the site $m$, the ground state  $|\tilde G_{p+is}\rangle$ can be written as
\begin{flalign}
|\tilde G_{p+is}\rangle=\frac{\varphi_{1,m-1}}{\sqrt{2}^4}\begin{pmatrix}\Lambda_{11}^{m}&\Lambda_{12}^{m}\\\Lambda_{21}^{m}&\Lambda_{22}^{m}\end{pmatrix}\begin{pmatrix}\Lambda_{11}^{m+1}&\Lambda_{12}^{m+1}\\\Lambda_{21}^{m+1}&\Lambda_{22}^{m+1}\end{pmatrix}|\varphi_{m+1,N}\rangle.\nn\\
\end{flalign}
The result shows magnetization vanishes in the bulk
\begin{eqnarray}
&&\langle \tilde G_{p+is}|S_m^z|\tilde G_{p+is}\rangle \qquad(1<m<N)\nn\\
&=&\frac{1}{16}\langle\varphi_{m+1,N}|
\begin{pmatrix}\Lambda_{11}^{m+1\dagger}&\Lambda_{21}^{m+1\dagger}\\\Lambda_{12}^{m+1\dagger}&\Lambda_{22}^{m+1\dagger}\end{pmatrix}\begin{pmatrix}\Lambda_{11}^{m\dagger}&\Lambda_{21}^{m\dagger}\\\Lambda_{12}^{m\dagger}&\Lambda_{22}^{m\dagger}\end{pmatrix}\nn\\
&&S_m^z\begin{pmatrix}\Lambda_{11}^{m}&\Lambda_{12}^{m}\\\Lambda_{21}^{m}&\Lambda_{22}^{m}\end{pmatrix}\begin{pmatrix}\Lambda_{11}^{m+1}&\Lambda_{12}^{m+1}\\\Lambda_{21}^{m+1}&\Lambda_{22}^{m+1}\end{pmatrix}
|\varphi_{m+1,N}\rangle\nn\\
&=&\frac{1}{16}\langle\varphi_{m+1,N}|\begin{pmatrix}0&0\\0&0\end{pmatrix}|\varphi_{m+1,N}\rangle=0
\end{eqnarray}

Aside from the analytical approach, spin magnetizations can also be obtained numerically. 
The numerical results for a chain of $N=20$ sites are shown in Figs. \ref{fig:edgemag} and \ref{fig:edge}.
The magnetizations are confined solely to the edges and vanish in the bulk. 
These numerical results at the edges are in good agreement with the analytical expressions obtained earlier, which is $\pm\frac{1}{4}(1-\frac{\Delta_s}{\sqrt{\Delta_s^2+\Delta_p^2}})$.

As long as the spin-$\frac12$ $p$-wave superconducting chain is in the same phase (see Appendix \ref{sec:0. topo_properties} for the phase diagram), the inclusion of an additional $s$-wave component will lead to similar behaviors,
though the restriction of the spin magnetization solely to the boundary sites only applies to the special cases $(\mu=0,\Delta_p=\pm t)$.
For more general situations, the magnetization is distributed over several sites near the edges and gradually decays to zero in the bulk.
Although the edge magnetizations can no longer be calculated using many-body wave functions,
they can be obtained by numerical calculations. 
Figure \ref{fig:variety} shows the distribution of the spin magnetizations for $(\mu=0,\Delta_p=5t,\Delta_s=t)$, $(\mu=t,\Delta_p=5t,\Delta_s=t)$, and $(\mu=1.7t,\Delta_p=5t,\Delta_s=t)$,
which confirms the distribution of magnetizations beyond the boundary sites.
In addition, by summing the magnetization over this edge region, the total edge magnetization at the two edges evolves from $\pm\frac14$ to $0$ with increasing strength of the $s$-wave component, as shown in Fig.\ref{fig:varymag}.

%For $N=40$, the summed magnetization for $(\mu=t,\Delta=\pm5t,\Delta_p=\Delta_s)$ deviates from the quantized value $\pm\frac14$ in the limit $\Delta_s/\Delta_p\to0$ due to finite-size effects. Increasing the system size to $N=80$ significantly suppresses the finite-size correction, and the summed magnetization approaches $\pm\frac14$ as $\Delta_s/\Delta_p\to0$.

%%%%%%%%%%%%%%%%%%%%%%%%%%%%%%%%%%%%%%%%%%%%%%%%%%%%%%%%%%%
\subsection{Local Majorana picture of edge spin}
\label{sec:local_majorana}

We present an alternative derivation of the edge spin magnetization in terms of the local Majorana operators shown in Fig. \ref{fig:chain}.
The advantage of this approach is that it allows one to clearly separate the contributions of boundary modes and scattering states to the edge magnetization.

The eigen operators of a spin-$\frac12$ $p+is$ superconductor contain both boundary eigen-modes and scattering eigen-modes.
While the boundary modes are bound states in the spectrum having wave function localized in the vicinity of the boundaries,
the scattering modes are extended throughout the whole chain. 
Reference \onlinecite{Yang2020} considers a 1D $p+is$ superconductor in the continuum limit with the Hamiltonian defined in Eq. (\ref{1D}),
in which it is found that the edge modes carry spin magnetizations $\pm \frac14$.
We have revealed that at the special parameters $(\mu=0,\Delta_p=\pm t)$,
the edge magnetization is given in Eq. (\ref{eq:magnetization_exact}),
which differs from $\pm\frac14$ by 
\begin{eqnarray}
M_{\text{scat}}=\pm\frac{\Delta_s}{4\sqrt{\Delta_s^2+\Delta_p^2}}.
\label{eq:M_scatter}
\end{eqnarray} 
This difference lies in the fact that Ref. \onlinecite{Yang2020} does not take into account the contributions from the scattering modes.

In the general cases, the scattering modes have their wave functions extended over the whole system,
as the case of the continuum model defined in Eq. (\ref{1D}).
On the other hand, the edge magnetizations are local quantities, which are localized around the boundary points within a range of localization length. 
The distinction between non localness of scattering modes and the localness of edge magnetizations makes the scattering mode contributions to the edge magnetizations not as intuitive as the boundary mode contributions. 

However, the parameters $(\mu=0,\Delta_p=\pm t)$ considered in this work have the advantage that the energy spectrum of the scattering modes is dispersionless, having the uniform value $\pm\frac{1}{2}\sqrt{\Delta_s^2+\Delta_p^2}$ as shown in Eq. (\ref{eq:H_exact_bulk}),
where the plus and minus signs correspond to quasi particle creation and annihilation operators, respectively. 
This dispersionless feature has an important consequence that the otherwise nonlocal scattering eigen modes can be properly combined into local ones  $\tilde\gamma_{B,x},\tilde\gamma_{A,x+1},\tilde{\gamma}_{B,x+1}',\tilde{\gamma}_{A,x}'$ defined in Eq. (\ref{eq:a_gamma_tilde}).
As a result, the contribution from scattering modes can be rendered much more transparent which we discuss below. 

Using Eqs. (\ref{eq:a_x_and_prime})and (\ref{eq:a_gamma}),
the local Majorana operators $\gamma_{A,x},\gamma_{B,x},\gamma_{A,x}',\gamma_{B,x}'$ are related to $c_{x\uparrow}^\dagger,c_{x\downarrow}^\dagger,c_{x\uparrow},c_{x\downarrow}$ via
\begin{eqnarray}
(\gamma_{A,x},\gamma_{B,x},\gamma_{A,x}',\gamma_{B,x}')=(c_{x\uparrow}^\dagger,c_{x\downarrow}^\dagger,c_{x\uparrow},c_{x\downarrow})U,
\end{eqnarray}
in which 
\begin{eqnarray}
U=\frac{1}{\sqrt{2}}
\begin{pmatrix}
i&1&i&1\\
i&1&-i&-1\\
-i&1&-i&1\\
-i&1&i&-1
\end{pmatrix}.
\end{eqnarray}
The local spin operators $S^z_x=\frac{1}{2}\left(c_{x\uparrow}^\dagger c_{x\uparrow}-c_{x\downarrow}^\dagger c_{x\downarrow}\right)$ ($1\leq x\leq N$)
can be written in terms of the local Majorana operators as 
\begin{eqnarray}
S^z_x=\frac{i}{4} (-\gamma_{A,x} \gamma^\prime_{B,x}+\gamma_{B,x}\gamma^\prime_{A,x}).
\label{eq:Szx_Majorana}
\end{eqnarray}
According to Fig. \ref{fig:chain}, when $x=1$, $\gamma_{A,1}$ and $\gamma^\prime_{B,1}$ in Eq. (\ref{eq:Szx_Majorana}) are boundary modes,
whereas $\gamma_{B,1}$ and $\gamma^\prime_{A,1}$ are scattering modes in the bulk, as they appear in the bulk Hamiltonian in Eq. (\ref{eq:H_ps_new_bulk}). 
Notice from Eq. (\ref{eq:H_ps_new}) that the edge Hamiltonian contains a term $\frac{i\Delta_s}{2}\gamma_{A,1} \gamma^\prime_{B,1}$,
which is minimized when $i\gamma_{A,1} \gamma^\prime_{B,1}=-1$.
Hence, the boundary mode contribution $-\frac{i}{4} \gamma_{A,1} \gamma^\prime_{B,1}$ to the ground-state value of $S^z_1$ is $\frac14$,
consistent with Ref. \onlinecite{Yang2020}. 

To evaluate the scattering mode contribution,
we rewrite $\frac{i}{4}\gamma_{B,1}\gamma^\prime_{A,1}$ in terms of $\tilde{\gamma}_{A,x},\tilde{\gamma}_{B,x},\tilde{\gamma}_{A,x}',\tilde{\gamma}_{B,x}'$ as 
\begin{eqnarray}
\frac{i}{4}\gamma_{B,1}\gamma^\prime_{A,1}&=& -\frac{i\Delta_s}{4\sqrt{\Delta_s^2+\Delta_p^2}}\tilde{\gamma}_{B,1}\tilde{\gamma}_{A,2}\nn\\
&&+\frac{i\Delta_p}{4\sqrt{\Delta_s^2+\Delta_p^2}}\tilde{\gamma}_{B,1}\tilde{\gamma}^\prime_{A,1}.
\label{eq:op_scattering_modes}
\end{eqnarray}
Recall from Sec. \ref{subsubsec:decouple_two_chains} that for the exact solution, the system consists of two decoupled chains as Fig. \ref{fig:chain3_1}, 
except that the operators in  $(\gamma_{A,1},\gamma_{B,1},\gamma_{A,1}',\gamma_{B,1}')$  in Fig. \ref{fig:chain3_1} have to be replaced with 
$(\tilde{\gamma}_{A,1},\tilde{\gamma}_{B,1},\tilde{\gamma}_{A,1}',\tilde{\gamma}_{B,1}')$.
Therefore, $i\tilde{\gamma}_{B,1}\tilde{\gamma}_{A,2}$ acquires the expectation value $-1$ in the ground state;
whereas the expectation value of $i\tilde{\gamma}_{B,1}\tilde{\gamma}^\prime_{A,1}$ is zero 
since $\tilde{\gamma}_{B,1}$ and $\tilde{\gamma}^\prime_{A,1}$ create quasi particle excitations in the upper and lower chains in Fig. \ref{fig:chain3_1}, respectively, which drive the system out of the ground-state subspace. 
This demonstrates that only the first term in Eq. (\ref{eq:op_scattering_modes})
has a non vanishing contribution to the ground-state expectation value of $\frac{i}{4}\gamma_{B,1}\gamma^\prime_{A,1}$. 
Hence, the scattering mode contribution to the spin magnetization is 
\begin{eqnarray}
\langle \frac{i}{4}\gamma_{B,1}\gamma^\prime_{A,1} \rangle=\frac{\Delta_s}{4\sqrt{\Delta_s^2+\Delta_p^2}},
\end{eqnarray}
which is exactly $M_{\text{scat}}$  in Eq. (\ref{eq:M_scatter}).

Finally, we also note that the local Majorana picture may have broader applications than edge spin magnetizations. 
It is known that $p+is$ superconductors have exotic response properties including axion electrodynamics, magneto electric effects,
and gravitational responses. 
The local Majorana picture at special sets of parameters may provide transparent and local pictures for these exotic response properties,
which will be helpful for establishing a better and more intuitive understanding of these phenomena
and will be left for future investigations.

%%%%%%%%%%%%%%%%%%%%%%%%%%%%%%%%%%%%%%%%%%%%%%%%%%%%%%%%%%%
%\vspace{0.8\baselineskip}
\section{Summary}
\label{sec:summary}

In summary, we have investigated the edge magnetization of the 1D spin-$\frac12$ $p\pm is$ superconductor using both analytical and numerical approaches. We show that opposite magnetizations emerge at the two ends of the system and evolve continuously with the strength of the $s$-wave pairing component. The edge magnetization is consistently supported by exact solutions and numerical calculations. 
Exact analytical ground-state wave functions are obtained at special parameter points. The system admits an effective description in terms of two coupled Kitaev spinless superconducting chains.

%%%%%%%%%%%%%%%%%%%%%%%%%%%%%%%%%%%%%%%%%%%
\begin{acknowledgments}

J.J. and W.Y. are supported by the Fundamental
Research Funds for the Central Universities.
C.X. is supported by the Shuimu Tsinghua Scholar program of Tsinghua University. 
C.W. is supported by the National Natural Science Foundation of China under the Grants No. 12234016 and No.12174317.
This work has been supported by the New Cornerstone Science Foundation.

\end{acknowledgments}

%%%%%%%%%%%%%%%%%%%%%%%%%%%%%%%%%%%%%%%%%%%%%%%%%%%%%%%%%%%%%%
%%%%%%%%%%%%%%%%%%%%%%%%%%%%%%%%%%%%%%%%%%%%%%%%%%%%%%%%%%%%%%
\appendix

\begin{widetext}
\section{Review of Kitaev model}

\subsection{Quasi particle annihilation and creation operators of Kitaev model}
\label{sec:-1. Quasi_operators}

From the Majorana operators $\gamma_{A,x}$ and $\gamma_{B,x}$, a set of quasi particle annihilation and creation operators $d_{x+1/2},d_{x+1/2}^\dagger$ can be defined on each bond (connecting sites $x$ and $x+1$) as
\begin{eqnarray}
d_{x+1/2}&=&\frac12\left(\gamma_{B,x}-i\gamma_{A,x+1}\right)=\frac12\left(-a_{x+1}+a_{x+1}^\dagger+a_x+a_x^\dagger\right), 
\end{eqnarray}
in which the site index $x$ runs from $1$ to $N-1$. 

For the case of $\Delta=t$, the Hamiltonian $\hat{H}_{K,+}$ can be further written as
\begin{eqnarray}
\hat{H}_{K,+}&=&t\sum_{x=1}^{N-1}\left(2d_{x+1/2}^\dagger d_{x+1/2}-1\right).
\end{eqnarray}
Notice in particular that the Hamiltonian does not contain $\gamma_{A,1}$ and $\gamma_{B,N}$,
which form zero-energy modes of the system. 
Since $\gamma_{A,1}$ and $\gamma_{B,N}$ can be recombined to a single complex fermion (with creation and annihilation operators being $\gamma_{A,1}\pm i\gamma_{B,N}$), the ground states are two fold degenerate.  
Since the states $(1\pm a_{x+1}^\dagger)(1\pm a_{x}^\dagger)|0\rangle$ are annihilated by the operator $d_{x+1/2}$, the two degenerate ground states $|\Psi^\pm(\Delta=t)\rangle$ can be obtained immediately as
\begin{eqnarray}
|\Psi^\pm(\Delta=t)\rangle=\prod_{x=1}^N\frac{1}{\sqrt2}\left(1\pm a_x^\dagger\right)|0\rangle, 
\end{eqnarray}

For the Hamiltonian $\hat{H}^{\text{D}}_p$, by defining $d_x^\dagger$ and $d_x^{\prime\dagger}$ as
\begin{eqnarray}
d_{x+1/2}^\dagger&=&\frac12\left(a_{x+1}+a_{x+1}^\dagger+a_{x}-a_{x}^\dagger\right),\nn\\
d_{x+1/2}^{\prime\dagger}&=&\frac12\left(a_{x+1}'+a_{x+1}'^\dagger+a_{x}'-a_{x}'^\dagger\right),
\end{eqnarray}
the Hamiltonian $\hat{H}^{\text{D}}_p$ can be rewritten as
\begin{eqnarray}
\hat{H}^{\text{D}}_p=2t\sum_{x=1}^{N-1}\left(d_{x+1/2}^\dagger d_{x+1/2}+d_{x+1/2}'^\dagger d'_{x+1/2}-1\right).
\end{eqnarray}
Therefore, the system has a flat excitation spectrum  
since the energy is raised by a uniform amount of $2t$ by applying $d_{x+1/2}^\dagger$ or $d_{x+1/2}'^\dagger$ for any $x$ ($1\leq x\leq N-1$).

\subsection{Topological properties}
\label{sec:0. topo_properties}

The topological properties of the Kitaev chain can be characterized by a winding number for the mapping from $k$ space to a two-dimensional parameter space. 

For the discussion of topological winding number, we consider PBC, such that the momentum $k_x$ is a good quantum number. 
The Hamiltonian $H(k_x)$ in momentum space for Kitaev's superconducting model in Eq.~\ref{Kitaev} is given by
\begin{eqnarray}
H_K\left(k_x\right)&=&\left(-2t\cos k_x-\mu\right)\tau_z-2\Delta\sin k_x\tau_y\nn\\
&=&h_z\left(k_x\right)\tau_z+h_y\left(k_x\right)\tau_y,
\label{eq:H_K_kx}
\end{eqnarray}
in which $\tau_y$ and $\tau_z$ are the Pauli operators in the particle-hole space. 
The real-valued vector $\vec h(k_x)=\left(h_y(k_x),h_z(k_x)\right)$ 
defines a mapping from the momentum space to two-component vectors. 
To characterize the topological property of this mapping, we define a complex function $q(k_x)$ as
\begin{eqnarray}
q\left(k_x\right)=\frac{h_z\left(k_x\right)+ih_y\left(k_x\right)}{|\vec h\left(k_x\right)|}.
\end{eqnarray}
As $k$ varies from $0$ to $2\pi$, the complex function $q(k_x)$ traces out a closed loop in the complex plane. The winding number $\omega$ is defined as the number of times this loop winds around the origin. 
More explicitly, $\omega$ is given by 
\begin{eqnarray}
\omega=\frac{1}{2\pi}\int_0^{2\pi}\frac{\mathrm d\arg\left[q\left(k_x\right)\right]}{\mathrm dk_x}\mathrm dk_x.
\end{eqnarray}

The winding number is an integer-valued topological invariant that can be used to distinguish different topological phases of the Kitaev superconducting chain. 
The phase diagram of the Kitaev's model is shown in Fig.\ref{fig:phase}.
The parameter choice $\mu=0,t>0,\Delta=t$ lies in the topological phase with winding number $\omega=+1$, while $\Delta=-t$ lies in a distinct topological phase with the winding number $\omega=-1$. 
Both of these two topological phases support unpaired Majorana fermions on the edges of the chain, 
though with different unpaired Majorana operators. 
Because of the robustness of the topological properties,
even if the parameters are away from the above-mentioned special points,
one still expects the emergence of one Majorana zero mode at each edge, as long as the winding number remains to be $1$ or $-1$.
In contrast, the $\omega=0$ case corresponds to the trivial phase, in which no unpaired Majorana zero modes exist at the edges.

%-------------------------------------------------------------------------
\begin{figure}
\includegraphics[width=0.45\textwidth]{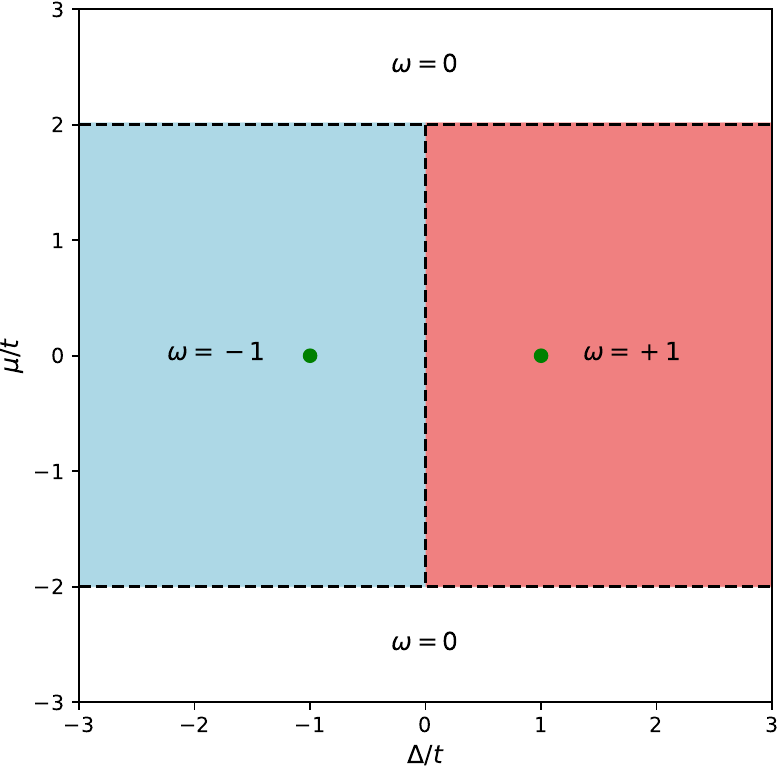}
\caption{The phase diagram of the Kitaev's model in the $(\Delta/t,\mu/t)$ plane. The topologically trivial regions $(\omega=0)$ occur for $|\mu|>2|t|$ (white). The system is topologically nontrivial for small chemical potential, $|\mu|<2|t|$. The two distinct topological phases are determined by the sign of $\Delta/t$: $\omega=+1$ for $\Delta/t>0$ (red) and $\omega=-1$ for $\Delta/t<0$ (blue). Two points with special parameter choices are marked in green within the topological regions.}
\label{fig:phase}
\end{figure}
%-------------------------------------------------------------------------

It is worth to note the continuum limit of the lattice model when the Fermi level lies close to the band bottom. 
The hopping term $t$ in Eq. (\ref{Kitaev}) gives a band dispersion of the form $-2t\cos(k_x)$,
which has band bottom at $k_x=0$ with energy $-2t$.
When chemical potential is slightly above $-2t$, the Fermi wave vector $k_F=\arccos(\frac{\mu}{2t})$ is very small,
hence,only small wave vectors are important for low-energy physics.  
An expansion of the band dispersion $-2t\cos(k_x)$ around $-2t$ gives a parabolic dispersion $-2t+t k_x^2+O(k_x^4)$,
with effective mass $m=1/(2t)$.
Therefore, in the weak pairing limit $\Delta\ll \frac{k_F^2}{2m}$, the lattice model in Eqs. (\ref{Kitaev})and(\ref{eq:H_K_kx}) reduces to the following continuum model:
\begin{eqnarray}
H_{K,\text{cont}}=\frac{1}{2m}(k_x^2-k_F^2)\tau_z-2\Delta k_x \tau_y.
\label{eq:H_K_cont}
\end{eqnarray}
When $\Delta>0$ and $\Delta<0$, the system described by $H_{K,\text{cont}}$ in Eq. (\ref{eq:H_K_cont}) resides at the lower left corner of the $\omega=1$ phase and lower right corner of the $\omega=-1$ phase, respectively. 
Therefore, as can be inspected from Fig. \ref{fig:phase},
 the continuum models with $\Delta>0$ and $\Delta<0$ are in the same topological phase as the special points $(\Delta=t,\mu=0)$ and $(\Delta=-t,\mu=0)$, respectively.
As a result, if one is only interested in topological properties,
it is legitimate to use $(\Delta=t,\mu=0)$ [or $(\Delta=-t,\mu=0)$] for the continuum model with $\Delta>0$ (or $\Delta<0$).

\section{The Summary of relations of operators}
\label{sec:I. The Summary of relations of operators}

The relations between $a_x$ and $c_x$ are given by
\begin{eqnarray}
a_x&=&\frac{1}{\sqrt{2}}\left(c_{x,\uparrow}+c_{x,\downarrow}\right),\nn\\
a_x'&=&\frac{1}{\sqrt{2}}\left(c_{x,\uparrow}-c_{x,\downarrow}\right),\nn\\
c_{x,\uparrow}&=&\frac{1}{\sqrt{2}}\left(a_{x}+a_{x}'\right),\nn\\
c_{x,\downarrow}&=&\frac{1}{\sqrt{2}}\left(a_{x}-a_{x}'\right).
\end{eqnarray}

The relations between $\gamma_x$ and $a_x$ are given by
\begin{eqnarray}
\gamma_{A,x}&=&-i\left(a_x-a_x^\dagger\right),\nn\\
\gamma_{B,x}&=&a_x+a_x^\dagger,\nn\\
a_x&=&\frac12\left(i\gamma_{A,x}+\gamma_{B,x}\right),\nn\\
a_x^\dagger&=&\frac12\left(-i\gamma_{A,x}+\gamma_{B,x}\right).
\end{eqnarray}

The relations between $\gamma_x$ and $c_x$ can be expressed in a matrix form
\begin{eqnarray}
\begin{pmatrix}
\gamma_{A,x}\\\gamma_{B,x}\\\gamma_{A,x}'\\\gamma_{B,x}'
\end{pmatrix}=\frac{1}{\sqrt{2}}
\begin{pmatrix}
-i&-i&i&i\\
1&1&1&1\\
-i&i&i&-i\\
1&-1&1&-1
\end{pmatrix}
\begin{pmatrix}
c_{x,\uparrow}\\c_{x,\downarrow}\\c_{x,\uparrow}^\dagger\\c_{x,\downarrow}^\dagger
\end{pmatrix}.
\end{eqnarray}

The action of $\gamma_{A,1}$ and $\gamma_{B,N}$ on $|\Psi^\pm(\Delta=t)\rangle$ can be derived as follows.
For $\gamma_{A,1}$, we have
\begin{eqnarray}
\gamma_{A,1}|\Psi^\pm(\Delta=t)\rangle&=&-i\left(a_1-a_1^\dagger\right)\frac{1}{\sqrt{2}^{N}}\left(1\pm a_N^\dagger\right)\dots\left(1\pm a_1^\dagger\right)|0\rangle\nn\\
&=&-i\frac{1}{\sqrt{2}^{N}}\left(1\mp a_N^\dagger\right)\dots\left(1\mp  a_2^\dagger\right)\left(a_1-a_1^\dagger\right)\left(1\pm a_1^\dagger\right)|0\rangle\nn\\
&=&\mp i\frac{1}{\sqrt{2}^{N}}\left(1\mp a_N^\dagger\right)\dots\left(1\mp a_2^\dagger\right)\left(1\mp a_1^\dagger\right)|0\rangle\nn\\
&=&\mp i |\Psi^\mp(\Delta=t)\rangle.
\end{eqnarray}
Similarly, for $\gamma_{B,N}$, we have
\begin{eqnarray}
\gamma_{B,N}|\Psi^\pm(\Delta=t)\rangle&=&\left(a_N+a_N^\dagger\right)\frac{1}{\sqrt{2}^{N}}\left(1\pm a_N^\dagger\right)\dots\left(1\pm a_1^\dagger\right)|0\rangle\nn\\
&=&\pm\frac{1}{\sqrt{2}^{N}}\left(1\pm a_N^\dagger\right)\dots\left(1\pm a_1^\dagger\right)|0\rangle\nn\\
&=&\pm  |\Psi^\pm(\Delta=t)\rangle.
\end{eqnarray}
The action of $\gamma_{A,N}$ and $\gamma_{B,1}$ on $|\Psi^\pm(\Delta=-t)\rangle$ is given by
\begin{eqnarray}
\gamma_{A,N}|\Psi^\pm(\Delta=-t)\rangle&=&-i\left(a_N-a_N^\dagger\right)\frac{1}{\sqrt{2}^{N}}\left(1\pm a_1^\dagger\right)\dots\left(1\pm a_N^\dagger\right)|0\rangle\nn\\
&=&-i\frac{1}{\sqrt{2}^{N}}\left(1\mp a_1^\dagger\right)\dots\left(1\mp  a_{N-1}^\dagger\right)\left(a_N-a_N^\dagger\right)\left(1\pm a_N^\dagger\right)|0\rangle\nn\\
&=&\mp i\frac{1}{\sqrt{2}^{N}}\left(1\mp a_1^\dagger\right)\dots\left(1\mp a_{N-1}^\dagger\right)\left(1\mp a_N^\dagger\right)|0\rangle\nn\\
&=&\mp i |\Psi^\mp(\Delta=-t)\rangle,\nn\\
\gamma_{B,1}|\Psi^\pm(\Delta=-t)\rangle&=&\left(a_1+a_1^\dagger\right)\frac{1}{\sqrt{2}^{N}}\left(1\pm a_1^\dagger\right)\dots\left(1\pm a_N^\dagger\right)|0\rangle\nn\\
&=&\pm\frac{1}{\sqrt{2}^{N}}\left(1\pm a_1^\dagger\right)\dots\left(1\pm a_N^\dagger\right)|0\rangle\nn\\
&=&\pm  |\Psi^\pm(\Delta=-t)\rangle.
\end{eqnarray}
Thus, we obtain the following relations:
\begin{eqnarray}
\gamma_{A,1}|\Psi^\pm(\Delta=t)\rangle&=&\mp i |\Psi^\mp(\Delta=t)\rangle,\nn\\
\gamma_{B,N}|\Psi^\pm(\Delta=t)\rangle&=&\pm |\Psi^\pm(\Delta=t)\rangle,\nn\\
\gamma_{A,N}|\Psi^\pm(\Delta=-t)\rangle&=&\mp i |\Psi^\mp(\Delta=-t)\rangle,\nn\\
\gamma_{B,1}|\Psi^\pm(\Delta=-t)\rangle&=&\pm |\Psi^\pm(\Delta=-t)\rangle 
\label{eq:gamma_act}
\end{eqnarray}

The quasiparticle annihilation operators of $|\Psi^{\pm\pm}\rangle$ satisfy 
\begin{eqnarray}
d_{x+1/2}&=&\frac12\left(\gamma_{B,x}-i\gamma_{A,x+1}\right)=\frac12\left(a_x+a_x^\dagger-a_{x+1}+a_{x+1}^\dagger\right),\nn\\
d_{x+1/2}^\dagger&=&\frac12\left(\gamma_{B,x}+i\gamma_{A,x+1}\right)=\frac12\left(a_x+a_x^\dagger+a_{x+1}-a_{x+1}^\dagger\right),\nn\\
d_{x+1/2}'&=&\frac12\left(\gamma_{B,x+1}'-i\gamma_{A,x}'\right)=\frac12\left(a_{x+1}'+a_{x+1}'^\dagger-a_{x}'+a_{x}'^\dagger\right),\nn\\
d_{x+1/2}'^\dagger&=&\frac12\left(\gamma_{B,x+1}'+i\gamma_{A,x}'\right)=\frac12\left(a_{x+1}'+a_{x+1}'^\dagger+a_{x}'-a_{x}'^\dagger\right),
\end{eqnarray}
and
\begin{eqnarray}
\gamma_{A,x+1}&=&-i\left(d_{x+1/2}^\dagger-d_{x+1/2}\right),\nn\\
\gamma_{B,x}&=&d_{x+1/2}+d_{x+1/2}^\dagger,\nn\\
\gamma_{A,x}'&=&-i\left(d_{x+1/2}'^\dagger-d_{x+1/2}'\right),\nn\\
\gamma_{B,x+1}'&=&d_{x+1/2}'+d_{x+1/2}'^\dagger.
\end{eqnarray}
These operators satisfy the fermionic anticommutation relations
\begin{eqnarray}
\{a_i,a_j^\dagger\}&=&\delta_{ij},\nn\\
\{\gamma_i,\gamma_j^\dagger\}&=&2\delta_{ij},\nn\\
\{d_i,d_j^\dagger\}&=&\delta_{ij}.
\end{eqnarray}

The transformation of $\gamma$ and $\tilde\gamma$ is given by
\begin{eqnarray}
\begin{pmatrix}\gamma_{B,x}\\\gamma_{A,x+1}\\\gamma_{A,x}'\\\gamma_{B,x+1}'\end{pmatrix}&=\begin{pmatrix}
1&0&0&0\\
0&\frac{\Delta_p}{\sqrt{\Delta_s^2+\Delta_p^2}}&\frac{\Delta_s}{\sqrt{\Delta_s^2+\Delta_p^2}}&0\\
0&-\frac{\Delta_s}{\sqrt{\Delta_s^2+\Delta_p^2}}&\frac{\Delta_p}{\sqrt{\Delta_s^2+\Delta_p^2}}&0\\
0&0&0&1
\end{pmatrix}\begin{pmatrix}\tilde\gamma_{B,x}\\\tilde\gamma_{A,x+1}\\\tilde\gamma_{A,x}'\\\tilde\gamma_{B,x+1}'\end{pmatrix}.
\end{eqnarray}
The relations between $\tilde a_x$ and $a_x$ are given by 
\begin{eqnarray}
\tilde a_1^\dagger&=&a_1^\dagger,\nn\\
\tilde a_x^\dagger&=&\frac12\left(-i\tilde\gamma_{A,x}+\tilde\gamma_{B,x}\right)\qquad(x\geq2)\nn\\
&=&\frac12\left[-\left(\frac{\Delta_p}{\sqrt{\Delta_s^2+\Delta_p^2}}\left(a_x-a_x^\dagger\right)-\frac{\Delta_s}{\sqrt{\Delta_s^2+\Delta_p^2}}\left(a'_{x-1}-a_{x-1}'^\dagger\right)\right)+a_x+a_x^\dagger\right],\nn\\
\tilde a_N'^\dagger&=&a_N'^\dagger,\nn\\
\tilde a_x'^\dagger&=&\frac12\left(-i\tilde\gamma'_{A,x}+\tilde\gamma'_{B,x}\right)\qquad(x\leq{N-1})\nn\\
&=&\frac12\left[-\left(\frac{\Delta_s}{\sqrt{\Delta_s^2+\Delta_p^2}}\left(a_{x+1}-a_{x+1}^\dagger\right)+\frac{\Delta_p}{\sqrt{\Delta_s^2+\Delta_p^2}}\left(a'_{x}-a_{x}'^\dagger\right)\right)+a'_x+a_x'^\dagger\right].
\end{eqnarray}
Then we can verify the relations $\tilde{a}_{m+1}\ket{\tilde{0}}=\tilde a'_{m}\ket{\tilde{0}}=0$.
We first rewrite $\ket{\tilde{0}}$ as
\begin{eqnarray}
\ket{\tilde{0}}=\left(u-v a_{m+1}^\dagger a_{m}'^\dagger\right)\ket{\tilde{0}_m}.
\end{eqnarray}
Notice that $\tilde{a}_{m+1}$ and $\tilde{a}'_{m}$ can be written in terms of $u$ and $v$:
\begin{eqnarray}
\tilde{a}_{m+1}&=&\frac{1}{2}\left[a_{m+1}\left(1+\frac{\Delta_p}{\sqrt{\Delta_s^2+\Delta_p^2}}\right)+a_{m+1}^\dagger\left(1-\frac{\Delta_p}{\sqrt{\Delta_s^2+\Delta_p^2}}\right)-a_{m}'\frac{\Delta_s}{\sqrt{\Delta_s^2+\Delta_p^2}}+a_m'^\dagger\frac{\Delta_s}{\sqrt{\Delta_s^2+\Delta_p^2}}\right]\nn\\
&=&\left(u^2 a_{m+1}+v^2 a_{m+1}^\dagger-uv a_{m}'+uv a_m'^\dagger\right),\nn\\
\tilde{a}'_{m}&=&\frac{1}{2}\left[a_{m+1}\frac{\Delta_s}{\sqrt{\Delta_s^2+\Delta_p^2}}-a_{m+1}^\dagger\frac{\Delta_s}{\sqrt{\Delta_s^2+\Delta_p^2}}+a_{m}'\left(1+\frac{\Delta_p}{\sqrt{\Delta_s^2+\Delta_p^2}}\right)+a_m'^\dagger\left(1-\frac{\Delta_p}{\sqrt{\Delta_s^2+\Delta_p^2}}\right)\right]\nn\\
&=&\left(uv a_{m+1}-uv a_{m+1}^\dagger+u^2 a_{m}'+v^2 a_m'^\dagger\right). 
\end{eqnarray}
Then we act $\tilde{a}_{m+1}$ and $\tilde{a}'_{m}$ on the vacuum state $\ket{\tilde{0}}$:
\begin{eqnarray}
\tilde{a}_{m+1}\ket{\tilde{0}}&=&\left(u^2 a_{m+1}+v^2 a_{m+1}^\dagger-uv a_{m}'+uv a_m'^\dagger\right)\left(u-v a_{m+1}^\dagger a_{m}'^\dagger\right)\ket{\tilde{0}_m}\nn\\
&=&\left(-u^2 v a_m'^\dagger+uv^2 a_{m+1}^\dagger-uv^2 a_{m+1}^\dagger+u^2 v a_m'^\dagger\right)\ket{\tilde{0}_m}\nn\\
&=&0,\nn\\
\tilde{a}'_{m}\ket{\tilde{0}}&=&\left(uv a_{m+1}-uv a_{m+1}^\dagger+u^2 a_{m}'+v^2 a_m'^\dagger\right)\left(u-v a_{m+1}^\dagger a_{m}'^\dagger\right)\ket{\tilde{0}_m}\nn\\
&=&\left(-uv^2 a_m'^\dagger-u^2v a_{m+1}^\dagger+u^2v a_{m+1}^\dagger+uv^2 a_m'^\dagger\right)\ket{\tilde{0}_m}\nn\\
&=&0.
\end{eqnarray}

\section{Degenerate perturbation for $p+is$ superconducting chain}
\label{sec:degenerate}

In this appendix, we treat the $s$-wave component in a perturbative manner. 
Exact solutions are discussed in Sec. \ref{sec:exact}.
The strategy is to treat the $s$-wave pairing using degenerate perturbation based on the equivalence between the 1D spin-$\frac12$ $p$-wave superconductor and two decoupled Kitaev spinless superconducting chains established in Sec. \ref{sec:topo}.
The many-body ground-state wave function under open boundary conditions will be obtained within degenerate perturbation,
and edge magnetizations are explicitly calculated from the obtained wave function. 
\subsection{Wave function in degenerate perturbation}

When $\Delta_s=0$, the red lines vanish and the system in Fig. \ref{fig:chain} reduces to two decoupled Kitaev superconducting chains as expected. 
Since the excitation spectrum of $\hat{H}^{\text{D}}_p$ is gapped, we can apply degenerate perturbation to the $s$-wave component. 
The inclusion of the $s$-wave component will lift the four fold degeneracy of the ground states in the $\Delta_s=0$ case.

Degenerate perturbation theory requires evaluating the matrix element $\langle\Psi^{\pm\pm} |\hat{H}_s|\Psi^{\pm\pm}\rangle$. 
Separating the edge and bulk terms, the Hamiltonian $\hat{H}_s$ can be rewritten as
\begin{eqnarray}
\hat{H}_s&=&\hat{H}_{s,\text{edge}}+\hat{H}_{s,\text{bulk}},
\end{eqnarray}
in which 
\begin{eqnarray}
\hat{H}_{s,\text{edge}}&=&\frac{i\Delta_s}{2}\left(\gamma_{A,1}\gamma_{B,1}'+\gamma_{B,N}\gamma_{A,N}'\right),\nn\\
\hat{H}_{s,\text{bulk}}&=&\frac{i\Delta_s}{2}\sum_{x=1}^{N-1} (\gamma_{B,x}\gamma^\prime_{A,x}+\gamma_{A,x+1}\gamma^\prime_{B,x+1}).\qquad 
\end{eqnarray}
Notice that  since
\begin{eqnarray}
\hat{H}_{s,\text{bulk}}=\Delta_s\sum_{x=1}^{N-1}\left(d_{x+1/2}^\dagger d_{x+1/2}'^\dagger+d_{x+1/2}'d_{x+1/2}\right).\;\;
\end{eqnarray}
$\hat{H}_{s,\text{bulk}}$ maps ground states to excited states. 
As a result, $\langle\Psi^{\pm\pm} |\hat{H}_{s,\text{bulk}}|\Psi^{\pm\pm}\rangle=0$,
and it is enough to consider the $\hat{H}_{s,\text{edge}}$ term. 

Define two sets of Pauli matrices $\sigma_\alpha$, $\sigma^\prime_\beta$ ($\alpha,\beta=x,y,z$) acting on the four-dimensional  subspace spanned by $|\Psi^{\pm\pm}\rangle$, in accordance with
\begin{eqnarray}
\sigma_z |\Psi^{\lambda\mu}\rangle&=& \lambda |\Psi^{\lambda\mu}\rangle,\nn\\
\sigma^\prime_z |\Psi^{\lambda\mu}\rangle&=& \mu |\Psi^{\lambda\mu}\rangle,
\end{eqnarray}
where $\lambda,\mu=\pm$.
Using Eq. (\ref{eq:gamma_act}),
the actions of $\gamma_{A,1}$, $\gamma_{B,N}$, $\gamma^\prime_{A,N}$, and $\gamma^\prime_{B,1}$ can be derived as $-\sigma_y$, $\sigma_z$, 
$-\sigma_x\sigma^\prime_y$, and $\sigma_x\sigma^\prime_z$, respectively,
in which the $\sigma_x$ factors in $\gamma_{A,N}$ and $\gamma_{B,1}$ are due to sign flips in $(1\pm a_x^\dagger)$'s when commuting $\gamma^\prime_{A,N},\gamma^\prime_{B,1}$ through $\prod_{x=1}^N\left(1\pm a_x^\dagger\right)$.
As a result, the restriction of $\hat{H}_{s,\text{edge}}$ in the four-dimensional subspace spanned by $|\Psi^{\pm\pm}\rangle$ becomes a $4\times 4$ matrix $H_{s,\text{res}}$, given by 
\begin{eqnarray}
H_{s,\text{res}}=\frac{\Delta_s}{2} (\sigma_y\sigma^\prime_y-\sigma_z\sigma^\prime_z).
\label{matrix}
\end{eqnarray}

The ground state of $H_{s,\text{res}}$ can be easily solved as $\frac{1}{\sqrt{2}}(1,0,0,1)^T$, with energy $-\frac{\Delta_s}{2}$.
Hence, the ground state of the 1D $p+is$ superconductor under OBC is non degenerate, which, in degenerate perturbation theory, is given by 
\begin{eqnarray}
|G_{p+is}\rangle=\frac{1}{\sqrt{2}}\left(|\Psi^{++}\rangle+|\Psi^{--}\rangle\right),
\end{eqnarray}
in which $|\Psi^{++}\rangle$ and $|\Psi^{--}\rangle$  are given  in Eq. (\ref{eq:Psi_pmpm}). 

%------------------------------------------------------------------------------------------------------------------------------------------------------------
\subsection{Edge magnetization in degenerate perturbation}
\label{subsec:edge_magnetization_approx}

Using the ground-state wave function obtained from degenerate perturbation theory, 
we next evaluate spin magnetizations on the edges. 

The spin operator at site $m$ is given by
\begin{eqnarray}
S_m^z&=&\frac12\left(c_{m,\uparrow}^\dagger c_{m,\uparrow}-c_{m,\downarrow}^\dagger c_{m,\downarrow}\right)
=\frac12\left(a_m^\dagger a_m'+a_m'^\dagger a_m\right),
\end{eqnarray}
where $a_m,a_m^\prime$ are defined in Eq. (\ref{eq:a_x_and_prime}). 
It is straightforward to verify that, for $m \neq m'$, the spin operator $S_m^z$ commutes with $(1 \pm a_{m'}^{\dagger}).$

For notational convenience, we introduce the symbols
$\phi_{n,m}^{\lambda}$ ($\lambda=\pm$ and $1\leq m<n\leq  N$) as 
\begin{eqnarray}
\phi_{n,m}^{\lambda}&=& \frac{1}{\sqrt{2}^{n-m+1}}(1+\lambda a^\dagger_n)(1+\lambda a^\dagger_{n-1})...(1+\lambda a^\dagger_m),\nn\\
\end{eqnarray}
and $\phi_{m,n}^{\prime\lambda}$ ($\lambda=\pm$ and $1\leq m<n\leq N$) as 
\begin{flalign}
\phi_{m,n}^{\prime\lambda}= \frac{1}{\sqrt{2}^{n-m+1}}(1+\lambda a^{\prime\dagger}_m)(1+\lambda a^{\prime\dagger}_{m+1})...(1+\lambda a^{\prime\dagger}_n).
\end{flalign}
Then $|\Psi^{++}\rangle$ and $|\Psi^{--}\rangle$ can be rearranged  as
\begin{eqnarray}
|\Psi^{++}\rangle%&=&\frac{1}{2}(1+a_{N}^\dagger)\psi^{++}_{N-1}(1+a_{N}'^\dagger)|0\rangle\nn\\
&=&\frac{1}{2}\big[(1+a_{N}^\dagger) \phi^+_{N-1,1} \phi^{\prime +}_{1,N-1}
+(1+a_{N}^\dagger)a_{N}'^\dagger\phi^-_{N-1,1} \phi^{\prime -}_{1,N-1}\big] |0\rangle,\nn\\
|\Psi^{--}\rangle%&=&\frac{1}{2}(1-a_{N}^\dagger)\psi^{--}_{N-1}(1-a_{N}'^\dagger)|0\rangle\nn\\
&=&\frac{1}{2}\big[(1-a_{N}^\dagger)\phi^-_{N-1,1} \phi^{\prime -}_{1,N-1}
-(1-a_{N}^\dagger)a_{N}'^\dagger\phi^+_{N-1,1} \phi^{\prime +}_{1,N-1}\big]|0\rangle.
\end{eqnarray}
The ground-state wave function $|G_{p+is}\rangle$ consequently  becomes
\begin{flalign}
&|G_{p+is}\rangle
=\frac{1}{2\sqrt{2}}[(1+a_{N}^\dagger-a_{N}'^\dagger+a_{N}^\dagger a_{N}'^\dagger)\phi^+_{N-1,1} \phi^{\prime +}_{1,N-1}
+(1-a_{N}^\dagger+a_{N}'^\dagger+a_{N}^\dagger a_{N}'^\dagger)\phi^-_{N-1,1} \phi^{\prime -}_{1,N-1}]|0\rangle.
\end{flalign}

Alternatively, we can express the ground state in terms of electron creation and annihilation operators as
\begin{flalign}
&|G_{p+is}\rangle=\frac{1}{2\sqrt{2}}[(1+\sqrt{2}c_{N,\downarrow}^\dagger-c_{N,\uparrow}^\dagger c_{N,\downarrow}^\dagger)\phi^+_{N-1,1} \phi^{\prime +}_{1,N-1}
+(1-\sqrt{2}c_{N,\downarrow}^\dagger-c_{N,\uparrow}^\dagger c_{N,\downarrow}^\dagger)\phi^-_{N-1,1} \phi^{\prime -}_{1,N-1}]|0\rangle,
\end{flalign}
which makes the spin structure at site $N$ transparent. 
Clearly, only the $c_{N,\downarrow}^\dagger$ terms contribute a nonzero spin for $S_N^z$. 
Since the weight of $c_{N,\downarrow}^\dagger$ in $\frac12(1\pm\sqrt{2}c_{N,\downarrow}^\dagger-c_{N,\uparrow}^\dagger c_{N,\downarrow}^\dagger)$ is $\frac12$, the spin expectation value of $S_N^z$ over the ground state is equal to $-\frac14$. 
This discussion makes the origin of the edge spin to be transparent on a wave function level. 
At the opposite edge, namely site $1$, the magnetization has the opposite sign and takes the value $+\frac14$.
One can also evaluate the magnetization $\langle G_{p+is}|S_N^z|G_{p+is}\rangle$ by directly calculating the expectation value.

We also demonstrate that the spin magnetization is confined exclusively to the edges,
by showing that the magnetization vanishes in the bulk. 
By breaking the products at site $m$, $|\Psi^{++}\rangle$ and $|\Psi^{--}\rangle$ can be rearranged  as
\begin{eqnarray}
|\Psi^{++}\rangle
&=&\frac{1}{2} [\phi_{N,m+1}^+\phi_{m-1,1}^+\phi^{\prime+}_{1,m-1}\phi^{\prime+}_{m+1,N}\ket{0}
+a_m^\dagger\phi_{N,m+1}^-\phi_{m-1,1}^+\phi^{\prime+}_{1,m-1}\phi^{\prime+}_{m+1,N}\ket{0}\nn\\
&&+a_m^{\prime\dagger} \phi_{N,m+1}^-\phi_{m-1,1}^-\phi^{\prime-}_{1,m-1}\phi^{\prime+}_{m+1,N}\ket{0}
+a_m^\dagger a_m^{\prime\dagger} \phi_{N,m+1}^+\phi_{m-1,1}^-\phi^{\prime-}_{1,m-1}\phi^{\prime+}_{m+1,N}\ket{0}],\qquad
\end{eqnarray}
and 
\begin{eqnarray}
|\Psi^{--}\rangle&=&\frac{1}{2} [\phi_{N,m+1}^-\phi_{m-1,1}^-\phi^{\prime-}_{1,m-1}\phi^{\prime-}_{m+1,N}\ket{0}
-a_m^\dagger\phi_{N,m+1}^+\phi_{m-1,1}^-\phi^{\prime-}_{1,m-1}\phi^{\prime-}_{m+1,N}\ket{0}\nn\\
&&-a_m^{\prime\dagger} \phi_{N,m+1}^+\phi_{m-1,1}^+\phi^{\prime+}_{1,m-1}\phi^{\prime-}_{m+1,N}\ket{0}
+a_m^\dagger a_m^{\prime\dagger} \phi_{N,m+1}^-\phi_{m-1,1}^+\phi^{\prime+}_{1,m-1}\phi^{\prime-}_{m+1,N}\ket{0}].\qquad
\end{eqnarray}
Notice that for $1< m<n< N$, all the eight terms in $\ket{G_{p+is}}=\frac{1}{\sqrt{2}}(|\Psi^{++}\rangle+|\Psi^{--}\rangle)$ are linearly independent.
Since $a_m^\dagger$ and $a_m^{\prime\dagger}$ are both equal weight combinations of $c_{m,\uparrow}^\dagger$ and $c_{m,\downarrow}^\dagger$,
each of the eight terms in $\ket{G_{p+is}}$ has zero net spin $S_m^z$ at site $m$. 
Therefore, the ground state does not have any spin magnetization in the bulk for $1<m<N$,
at least within the approximation of degenerate perturbation theory for small $\Delta_s$.

It is clear why there is no scattering mode contribution to the edge magnetization within degenerate perturbation theory.
In degenerate perturbation, the wave function is the bulk is untouched,
namely, all the vertical red lines in Fig. \ref{fig:chain} are absent.
In this case, there is no mixing of Majorana operators in the bulk, and as a result, 
$\tilde{\gamma}_{B,1}= \gamma_{B,1}$, $\tilde{\gamma}^\prime_{A,1}=\gamma^\prime_{A,1}$. 
Unlike Eq. (\ref{eq:op_scattering_modes}), we have
$\frac{i}{4}\gamma_{B,1}\gamma^\prime_{A,1}\approx \frac{i}{4}\tilde{\gamma}_{B,1}\tilde{\gamma}^\prime_{A,1}$
in degenerate perturbation,
which has zero expectation value as discussed below Eq. (\ref{eq:op_scattering_modes}).

\section{The recursive form of the exact ground-state wave function}
\label{sec:III. The recursive form of the exact ground state wave function}

We first prove the state $|\tilde\Psi^{\pm\pm}\rangle$ can be written in the following recursive matrix product form. We consider the state $|\tilde\Psi^{++}\rangle$:
\begin{eqnarray}
|\tilde \Psi^{++}\rangle=\frac{1}{2^N}\begin{pmatrix}\left(1+\tilde a_1^\dagger\right)\left(1+\tilde a_1'^\dagger\right)&\left(1-\tilde a_1^\dagger\right)\left(1-\tilde a_1'^\dagger\right)\end{pmatrix}
\prod_{x=N}^2\begin{pmatrix}
1+\tilde a_x'^\dagger&-\tilde a_x^\dagger\left(1-\tilde a_x'^\dagger\right)\\
\tilde a_x^\dagger\left(1+\tilde a_x'^\dagger\right)&1-\tilde a_x'^\dagger
\end{pmatrix}\hat f_N^{++}|\tilde0\rangle. 
\label{eq:D1}
\end{eqnarray}
We assume that the state $|\tilde\Psi^{++}\rangle$ can be written as the following representation for arbitary $m>1$.
\begin{eqnarray}
|\tilde \Psi^{++}\rangle&=&\frac{1}{2^N}\left[\left(1+\tilde a_m^\dagger\right)\left(1+\tilde a_{m-1}^\dagger\right)\dots\left(1+\tilde a_1^\dagger\right)\left(1+\tilde a_1'^\dagger\right)\dots\left(1+\tilde a_{m-1}'^\dagger\right)\left(1+\tilde a_m'^\dagger\right)\hat h_{m}^{++}|\tilde0\rangle\right.\nn\\
&&+\left.\left(1-\tilde a_m^\dagger\right)\left(1-\tilde a_{m-1}^\dagger\right)\dots\left(1-\tilde a_1^\dagger\right)\left(1-\tilde a_1'^\dagger\right)\dots\left(1-\tilde a_{m-1}'^\dagger\right)\left(1-\tilde a_m'^\dagger\right)\hat h_{m}^{--}|\tilde0\rangle\right].
\end{eqnarray}
The state $|\tilde\Psi^{++}\rangle$ can be rewritten as
\begin{eqnarray}
|\tilde \Psi^{++}\rangle&=&\frac{1}{2^N}\left[\left(1+\tilde a_{m-1}^\dagger\right)\dots\left(1+\tilde a_1^\dagger\right)\left(1+\tilde a_1'^\dagger\right)\dots\left(1+\tilde a_{m-1}'^\dagger\right)\left(1+\tilde a_m'^\dagger\right)\hat h_{m}^{++}|\tilde0\rangle\right.\nn\\
&&+\left(1-\tilde a_{m-1}^\dagger\right)\dots\left(1-\tilde a_1^\dagger\right)\left(1-\tilde a_1'^\dagger\right)\dots\left(1-\tilde a_{m-1}'^\dagger\right)\tilde a_m^\dagger\left(1+\tilde a_m'^\dagger\right)\hat h_{m}^{++}|\tilde0\rangle\nn\\
&&+\left(1-\tilde a_m^\dagger\right)\left(1-\tilde a_{m-1}^\dagger\right)\dots\left(1-\tilde a_1^\dagger\right)\left(1-\tilde a_1'^\dagger\right)\dots\left(1-\tilde a_{m-1}'^\dagger\right)\left(1-\tilde a_m'^\dagger\right]\hat h_{m}^{--}|\tilde0\rangle\nn\\
&&-\left.\left(1+\tilde a_{m-1}^\dagger\right)\dots\left(1+\tilde a_1^\dagger\right)\left(1+\tilde a_1'^\dagger\right)\dots\left(1+\tilde a_{m-1}'^\dagger\right)\tilde a_m^\dagger\left(1-\tilde a_m'^\dagger\right)\hat h_{m}^{--}|\tilde0\rangle\right]\nn\\
&=&\frac{1}{2^N}\left[\left(1+\tilde a_{m-1}^\dagger\right)\dots\left(1+\tilde a_1^\dagger\right)\left(1+\tilde a_1'^\dagger\right)\dots\left(1+\tilde a_{m-1}'^\dagger\right)\hat h_{m-1}^{++}|\tilde0\rangle\right.\nn\\
&&+\left.\left(1-\tilde a_{m-1}^\dagger\right)\dots\left(1-\tilde a_1^\dagger\right)\left(1-\tilde a_1'^\dagger\right)\dots\left(1-\tilde a_{m-1}'^\dagger\right)\hat h_{m-1}^{--}|\tilde0\rangle\right].
\end{eqnarray}
The recursive relation between $\hat h_{m-1}$ and $\hat h_{m}$ can be derived as follows:
\begin{eqnarray}
\begin{pmatrix}\hat h_{m-1}^{++}\\\hat h_{m-1}^{--}\end{pmatrix}=
\begin{pmatrix}
1+\tilde a_m'^\dagger&-\tilde a_m^\dagger\left(1-\tilde a_m'^\dagger\right)\\
\tilde a_m^\dagger\left(1+\tilde a_m'^\dagger\right)&1-\tilde a_m'^\dagger
\end{pmatrix}
\begin{pmatrix}\hat h_{m}^{++}\\\hat h_{m}^{--}\end{pmatrix}.
\end{eqnarray}
For the state $|\tilde\Psi^{++}\rangle$, when we take $m = N$, it is clear that $\hat{h}_{N}^{++} = 1$ and $\hat{h}_{N}^{--} = 0$. Accordingly, the initial matrix of $|\tilde\Psi^{++}\rangle$ is given by $f_N^{++}=\begin{pmatrix}1&0\end{pmatrix}^T$. For the state $|\tilde\Psi^{--}\rangle$, the recursive relation is the same as the previous case, except that the initial matrix is different, which is $f_N^{--}=\begin{pmatrix}0&1\end{pmatrix}^T$.

We define the column vector $\tilde\eta_1$ for convenience
\begin{eqnarray}
\tilde\eta_1=\begin{pmatrix}\left(1+\tilde a_1^\dagger\right)\left(1+\tilde a_1'^\dagger\right),\left(1-\tilde a_1^\dagger\right)\left(1-\tilde a_1'^\dagger\right)\end{pmatrix}^{T}
\end{eqnarray}

After obtaining the recursive form in Eq.~(\ref{eq:D1}), we factor out the operators $u-v a_{x}^\dagger a_{x-1}'^\dagger$ from the state $|\tilde{0}\rangle$ to derive an expression without $\tilde a^\dagger$ and $\tilde a'^\dagger$. For the state $|\tilde\Psi^{++}\rangle$, we first apply the matrix at $x=N$ on $\hat f_N^{++}$ and define the corresponding column vector $\hat g_N^{++}$ as
\begin{eqnarray}
\hat g_N^{++}|0\rangle&=&\begin{pmatrix}\left(1+a_N'^\dagger\right)\hat h_N^{++}-\tilde a_N^\dagger\left(1-a_N'^\dagger\right)\hat h_N^{--}\\
\left(1-a_N'^\dagger\right)\hat h_N^{--}+\tilde a_N^\dagger\left(1+a_N'^\dagger\right)\hat h_N^{++}\end{pmatrix}|0\rangle=\begin{pmatrix}J_N+\tilde a_N^\dagger J_N'\\K_N+\tilde a_N^\dagger K_N'\end{pmatrix}|0\rangle.
\end{eqnarray}
The column vector $\hat g_m^{++}$ satisfies the recursive relation with $\hat g_{m-1}^{++}$:
\begin{eqnarray}
\hat g_{m-1}^{++}|0\rangle=\begin{pmatrix}
1+\tilde a_{m-1}'^\dagger&-\tilde a_{m-1}^\dagger\left(1-\tilde a_{m-1}'^\dagger\right)\\
\tilde a_{m-1}^\dagger\left(1+\tilde a_{m-1}'^\dagger\right)&1-\tilde a_{m-1}'^\dagger
\end{pmatrix}\hat g_{m}^{++}\left(u-va_{m}^\dagger a_{m-1}'^\dagger\right)|0\rangle,
\end{eqnarray}
\begin{flalign}
\begin{pmatrix}J_{m-1}+\tilde a_{m-1}^\dagger J_{m-1}'\\K_{m-1}+\tilde a_{m-1}^\dagger K_{m-1}'\end{pmatrix}|0\rangle=\begin{pmatrix}
\left(1+\tilde a_{m-1}'^\dagger\right)\left(J_{m}+\tilde a_{m}^\dagger J_{m}'\right)-\tilde a_{m-1}^\dagger\left(1-\tilde a_{m-1}'^\dagger\right)\left(K_{m}+\tilde a_{m}^\dagger K_{m}'\right)\\
\left(1-\tilde a_{m-1}'^\dagger\right)\left(K_{m}+\tilde a_{m}^\dagger K_{m}'\right)+\tilde a_{m-1}^\dagger\left(1+\tilde a_{m-1}'^\dagger\right)\left(J_{m}+\tilde a_{m}^\dagger J_{m}'\right)
\end{pmatrix}\left(u-va_{m}^\dagger a_{m-1}'^\dagger\right)|0\rangle,\nn\\
\end{flalign}
from which the recursion relations follow:
\begin{eqnarray}
J_{m-1}&=&\left(u\lambda_1^m+u\lambda_2^m+v\lambda_3^m-\lambda_4^m\right)J_m+\left(-v\lambda_1^m-v\lambda_2^m+u\lambda_3^m-u\lambda_4^m\right)J_m',\nn\\
J_{m-1}'&=&\left(-u\lambda_1^m+u\lambda_2^m+v\lambda_3^m+v\lambda_4^m\right)K_m+\left(-v\lambda_1^m+v\lambda_2^m-u\lambda_3^m-u\lambda_4^m\right)K_m',\nn\\
K_{m-1}&=&\left(u\lambda_1^m-u\lambda_2^m-v\lambda_3^m-v\lambda_4^m\right)K_m+\left(v\lambda_1^m-v\lambda_2^m+u\lambda_3^m+u\lambda_4^m\right)K_m'=-J_{m-1}',\nn\\
K_{m-1}'&=&\left(u\lambda_1^m+u\lambda_2^m+v\lambda_3^m-v\lambda_4^m\right)J_m+\left(-v\lambda_1^m-v\lambda_2^m+u\lambda_3^m-u\lambda_4^m\right)J_m'=J_{m-1},
\end{eqnarray}
in which 
\begin{eqnarray}
\lambda_1^m=1,\quad\lambda_2^m=a_{m-1}'^\dagger,\quad\lambda_3^m=a_{m}^\dagger,\quad\lambda_4^m=a_{m}^\dagger a_{m-1}'^\dagger.
\end{eqnarray}
Since for all $x$ we have $K_x' = J_x$ and $J_x' = -K_x$, the recursive relations can be simplified to the matrix product form that involves only $J_x$ and $K_x$.
Thus, the state $|\tilde\Psi^{++}\rangle$ is given by
\begin{eqnarray}
|\tilde \Psi^{++}\rangle&=&\frac{1}{2^N}\tilde\eta_1^T
\prod_{x=N}^2\begin{pmatrix}
1+\tilde a_x'^\dagger&-\tilde a_x^\dagger\left(1-\tilde a_x'^\dagger\right)\\
\tilde a_x^\dagger\left(1+\tilde a_x'^\dagger\right)&1-\tilde a_x'^\dagger
\end{pmatrix}\hat f_N^{++}|\tilde0\rangle\nn\\
&=&\frac{1}{2^N}\tilde\eta_1^T
\prod_{x=N}^2\left[\left(u+va_{x}^\dagger a_{x-1}'^\dagger\right)\begin{pmatrix}
1+\tilde a_x'^\dagger&-\tilde a_x^\dagger\left(1-\tilde a_x'^\dagger\right)\\
\tilde a_x^\dagger\left(1+\tilde a_x'^\dagger\right)&1-\tilde a_x'^\dagger
\end{pmatrix}\right]\begin{pmatrix}\hat h_{N}^{++}\\\hat h_{N}^{--}\end{pmatrix}|0\rangle\nn\\
&=&\frac{1}{2^N}\tilde\eta_1^T\left(u-va_{2}^\dagger a_{1}'^\dagger\right)
\prod_{x=N-1}^2\left[\begin{pmatrix}
1+\tilde a_x'^\dagger&-\tilde a_x^\dagger\left(1-\tilde a_x'^\dagger\right)\\
\tilde a_x^\dagger\left(1+\tilde a_x'^\dagger\right)&1-\tilde a_x'^\dagger
\end{pmatrix}\left(u-va_{x+1}^\dagger a_{x}'^\dagger\right)\right]\hat g_N^{++}|0\rangle\nn\\
&=&\frac{1}{2^N}\tilde\eta_1^T\left(u-va_{2}^\dagger a_{1}'^\dagger\right)
\prod_{x=m-1}^2\left[\begin{pmatrix}
1+\tilde a_x'^\dagger&-\tilde a_x^\dagger\left(1-\tilde a_x'^\dagger\right)\\
\tilde a_x^\dagger\left(1+\tilde a_x'^\dagger\right)&1-\tilde a_x'^\dagger
\end{pmatrix}\left(u-va_{x+1}^\dagger a_{x}'^\dagger\right)\right]\hat g_m^{++}|0\rangle\nn\\
&=&\frac{1}{2^N}\tilde\eta_1^T\left(u-va_{2}^\dagger a_{1}'^\dagger\right)
\left[\begin{pmatrix}
1+\tilde a_2'^\dagger&-\tilde a_2^\dagger\left(1-\tilde a_2'^\dagger\right)\\
\tilde a_2^\dagger\left(1+\tilde a_2'^\dagger\right)&1-\tilde a_2'^\dagger
\end{pmatrix}\left(u-va_{3}^\dagger a_{2}'^\dagger\right)\right]\hat g_3^{++}|0\rangle\nn\\
&=&\frac{1}{2^N}\tilde\eta_1^T\left(u-va_{2}^\dagger a_{1}'^\dagger\right)
\hat g_2^{++}|0\rangle\nn\\
&=&\frac{1}{2^N}\tilde\eta_1^T\begin{pmatrix}J_2-\tilde a_2^\dagger K_2\\K_2+\tilde a_2^\dagger J_2\end{pmatrix}\left(u-va_{2}^\dagger a_{1}'^\dagger\right)
|0\rangle.
\end{eqnarray}

By iteration, we obtain the explicit form of $\hat g_2^{++}$. For $3\leq m\leq N$, the recursion relations satisfy
\begin{eqnarray}
J_{m-1}|0\rangle&=&\left(1+\tilde a_{m-1}'^\dagger\right)\left(J_{m}-\tilde a_{m}^\dagger K_{m}\right)\left(u-va_{m}^\dagger a_{m-1}'^\dagger\right)|0\rangle,\nn\\
K_{m-1}|0\rangle&=&\left(1-\tilde a_{m-1}'^\dagger\right)\left(K_{m}+\tilde a_{m}^\dagger J_{m}\right)\left(u-va_{m}^\dagger a_{m-1}'^\dagger\right)|0\rangle,\nn
\end{eqnarray}
We can extend the definition to $J_1$ and $K_1$, and the state $|\tilde\Psi^{++}\rangle$ can be expressed as
\begin{eqnarray}
|\tilde\Psi^{++}\rangle&=&\frac{1}{2^N}
\left[\left(1+a_1^\dagger\right)\left(1+\tilde a_1'^\dagger\right)\left(J_2-\tilde a_2^\dagger K_2\right)+\left(1-a_1^\dagger\right)\left(1-\tilde a_1'^\dagger\right)\left(K_2+\tilde a_2^\dagger J_2\right)\right)\left(u-va_{2}^\dagger a_{1}'^\dagger\right]|0\rangle\nn\\
&=&\frac{1}{2^N}\left[\left(1+a_1^\dagger\right)J_1+\left(1-a_1^\dagger\right)K_1\right]|0\rangle.
\end{eqnarray}
For the ground state $|\tilde G_{p+is}\rangle$, the vector $\hat g_N^{++}$ is replaced by $\hat g_N$, with $\hat h_N^{++}=\hat h_N^{--}=\tfrac{1}{\sqrt{2}}$. We factor out the prefactor $\tfrac{1}{\sqrt{2}}$ to the front.
Finally, by expanding $J_1$ and $K_1$ in the above expression via the recursive relations back to the initial operators $J_N$ and $K_N$, the proof is completed.

\end{widetext}

%%%%%%%%%%%%%%%%%%%%%%%%%%%%%%%%%%%%%%%%%%%%%%%%%%%%%%%%%%%%%%%

\end{document}